\documentclass[journal]{IEEEtran}
\usepackage[compatibility=false]{caption}

\IEEEoverridecommandlockouts

\usepackage{cite}
\usepackage{hyperref}
\usepackage{tikz}
\usepackage{amsmath,amssymb,amsfonts,mathtools,bm}
\usepackage{amsthm}
\usepackage{graphicx}
\usepackage{textcomp}
\usepackage{xcolor}
\usepackage{booktabs}
\usepackage{multirow}
\usepackage{subcaption}
\usepackage{algorithm}
\usepackage{algpseudocode}
\usepackage{url}
\usepackage{balance}
\usepackage{pifont}
\usepackage{adjustbox} 

\definecolor{ourred}{RGB}{200,0,0} 
\newcommand{\ourmodelval}[1]{\textcolor{ourred}{\textbf{#1}}}

\definecolor{ourblue}{RGB}{0,0,150} 
\newcommand{\secondbestval}[1]{\textcolor{ourblue}{\underline{#1}}}

\definecolor{darkgreen}{HTML}{248a73} 

\DeclareRobustCommand{\orcidicon}{%
    \begin{tikzpicture}
    \draw[lime, fill=lime] (0,0) 
    circle [radius=0.16] 
    node[white] {{\fontfamily{qag}\selectfont \tiny ID}};     \draw[white, fill=white] (-0.0625,0.095) 
    circle [radius=0.007];     \end{tikzpicture}
    \hspace{-2mm}}
\foreach \x in {A, ..., Z}{%
    \expandafter\xdef\csname orcid\x\endcsname{\noexpand\href{https://orcid.org/\csname orcidauthor\x\endcsname}{\noexpand\orcidicon}}
    }

\allowdisplaybreaks
\renewcommand{\baselinestretch}{1}

\begin{document}

\title{RadioMamba: Breaking the Accuracy-Efficiency Trade-off in Radio Map Construction via a Hybrid Mamba-UNet}

\author{
Honggang Jia\orcidA{},~\IEEEmembership{Student Member,~IEEE,}
Nan Cheng\orcidB{},~\IEEEmembership{Senior Member,~IEEE,}\\
Xiucheng Wang\orcidC{},~\IEEEmembership{Student Member,~IEEE,}
Conghao Zhou\orcidD{},~\IEEEmembership{Member,~IEEE,}\\
Ruijin Sun\orcidE{},~\IEEEmembership{Member,~IEEE,}
Xuemin (Sherman) Shen\orcidF{},~\IEEEmembership{Fellow,~IEEE}


\thanks{ }
\thanks{
\par This work was supported by the National Key Research and Development Program of China (2024YFB2907500).
\par Honggang Jia, Nan Cheng, Xiucheng Wang, Conghao Zhou, Ruijin Sun are with the State Key Laboratory of ISN and School of Telecommunications Engineering, Xidian University, Xi’an 710071, China (e-mail: jiahg@stu.xidian.edu.cn; dr.nan.cheng@ieee.org; xcwang\_1@stu.xidian.edu.cn; conghao.zhou@ieee.org; sunruijin@xidian.edu.cn. \textit{Nan Cheng is the corresponding author}.
\par Xuemin (Sherman) Shen is with the Department of Electrical and Computer Engineering, University of Waterloo, Waterloo, N2L 3G1, Canada (e-mail: sshen@uwaterloo.ca).
}

}

\maketitle

\begin{abstract}
Radio map (RM) has recently attracted much attention since it can provide real-time and accurate spatial channel information for 6G services and applications. However, current deep learning-based methods for RM construction exhibit well known accuracy-efficiency trade-off. In this paper, we introduce RadioMamba, a hybrid Mamba-UNet architecture for RM construction to address the trade-off. Generally, accurate RM construction requires modeling long-range spatial dependencies, reflecting the global nature of wave propagation physics. RadioMamba utilizes a Mamba-Convolutional block where the Mamba branch captures these global dependencies with linear complexity, while a parallel convolutional branch extracts local features. This hybrid design generates feature representations that capture both global context and local detail. Experiments show that RadioMamba achieves higher accuracy than existing methods, including diffusion models, while operating nearly 20 times faster and using only 2.9\% of the model parameters. By improving both accuracy and efficiency, RadioMamba presents a viable approach for real-time intelligent optimization in next generation wireless systems.
\end{abstract}


\section{Introduction}
\label{sec:introduction}

\begin{figure}[ht]
    \centering
    \captionsetup{font=small}
    \includegraphics[width=0.8\columnwidth]{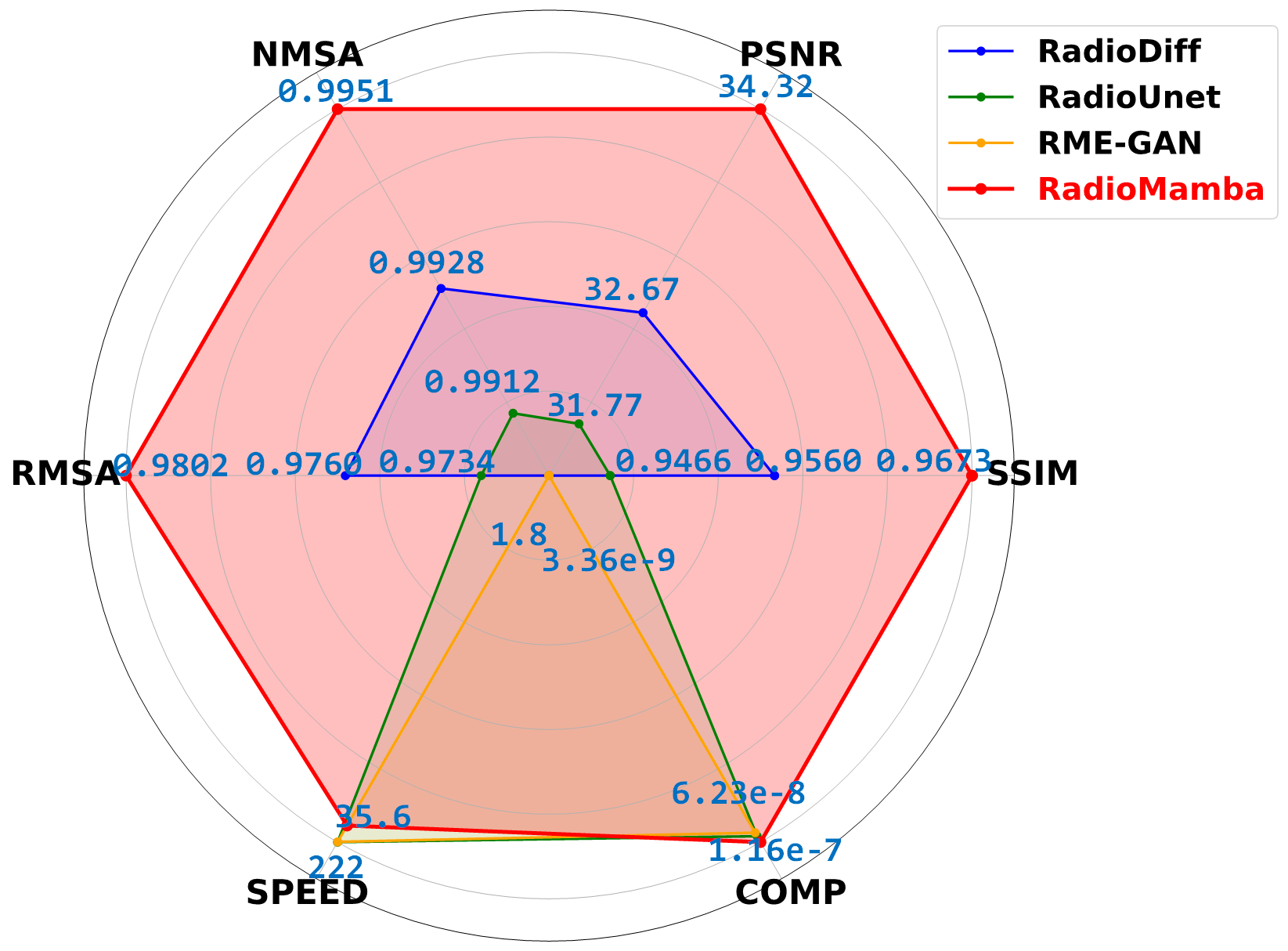}
    \caption{A radar chart comparing four RM construction models across six metrics: SSIM, PSNR, NMSA (calculated as 1-NMSE), RMSA (calculated as 1-RMSE), SPEED (inverse of inference time), and COMP (model compactness, inverse of parameter count).}
    \label{fig:radar_chart}
\end{figure}

The continuous advancement towards sixth-generation (6G) wireless networks is enabling a future with the internet of things (IoT), autonomous systems, and immersive cyber-physical experiences \cite{dang2020should, tao2022digital}. This vision, which underpins future applications like the metaverse \cite{1}, motivates the transition from reactive network management to proactive, data-driven control. A key component of this evolution is the network digital twin (NDT) \cite{lin20236g, 3}, a high-fidelity virtual replica of the physical network environment that enables simulation, prediction, and optimization in real-time \cite{wu2021digital, letaief2019roadmap, ansari2025toward}. The NDT concept is becoming a cornerstone for managing the unprecedented complexity of 6G systems, offering a paradigm to operate, manage, and optimize wireless networks with enhanced intelligence and autonomy \cite{3}.

For wireless systems, especially those involving mobile elements like UAVs \cite{13}, a high-fidelity twin must extend beyond the physical infrastructure to also model the complex radio propagation environment, a task that comes with its own set of security and reliability challenges \cite{2}. This functionality is provided by a radio map (RM)—a detailed spatial representation of signal propagation characteristics, such as pathloss or received signal strength\cite{morais2024localization}. The precision of such maps can be further enhanced by technologies like integrated sensing and communication (ISAC) \cite{liu2022integrated, liu2022survey}. Consequently, an accurate and continuously updated RM is essential for the NDT and a wide range of other critical 6G applications. These include dynamic spectrum management, interference coordination, user-cell association, and on-the-fly trajectory planning for unmanned aerial vehicles (UAVs) in complex air-ground networks \cite{bi2019engineering, zhang2020radio, zhang2023reconfigurable, 8}. Furthermore, high-definition maps, a specialized form of RMs, are critical for ensuring the safety and efficiency of vehicular networks by providing centimeter-level accuracy for autonomous vehicle navigation and coordination \cite{12}.

However, traditional RM construction methods have several limitations. These traditional approaches are broadly categorized into two types: measurement-based and physical model-based. Measurement-based techniques rely on deploying extensive sensors or conducting drive-tests to gather signal measurements, which are then used with interpolation algorithms like Kriging to construct the RM \cite{mao2018constructing, cover1967nearest, breidt2000local}. However, these methods are often expensive, labor-intensive, and face challenges in scaling in complex urban environments. The quality of the resulting RM is highly dependent on the quantity and accuracy of the measurements, and these methods are challenged in constructing maps for inaccessible regions, limiting their use in applications like UAV path planning\cite{gong20233d,dong2022radio, 11}. Alternatively, physical model-based methods using electromagnetic ray-tracing (ERT) \cite{oh2004mimo, salski2020electromagnetic} can yield highly precise RMs by simulating wave propagation in a 3D environmental model. Yet, they are computationally intensive. This high latency makes both traditional approaches less suitable for dynamic 6G scenarios where the environment, affected by elements like moving vehicles, changes in seconds \cite{tshakwanda2024advancing, nandy2025exploring}.

To address these challenges, the community has turned to deep learning (DL), which can learn the complex, non-linear mapping from environmental features to a complete RM with rapid inference capabilities. Initial works like RadioUNet \cite{levie2021radiounet} and RME-GAN \cite{zhang2023rme} established the viability of this approach. Most recently, diffusion models, exemplified by RadioDiff \cite{wang2024radiodiff}, have become the state-of-the-art (SOTA) in terms of raw accuracy, leveraging an iterative denoising process to generate high-fidelity maps. While their inference process is faster than traditional ray-tracing simulations, the iterative nature of diffusion models still incurs latencies that are too high for the most demanding 6G applications \cite{luo2025denoising}. Scenarios such as real-time network digital twins require near-instantaneous environmental updates, a requirement that current SOTA models cannot meet. This situation creates a trade-off: a choice between the high accuracy of heavyweight generative models and the low latency of their less accurate, lightweight predecessors. Overcoming this trade-off is critical for the widespread deployment of DL-based RM solutions in future real-time systems.

This accuracy-efficiency trade-off is not merely an issue of model scaling but is deeply rooted in the foundational architectures of current deep learning models, which are often adapted from the domain of computer vision. In radio map construction, the input is typically an image-like representation of the environment\cite{levie2021radiounet} where long-range physical phenomena, such as reflections and diffractions, manifest as subtle, spatially distant correlations. While convolutional neural networks (CNNs) are efficient, their hierarchical nature, built upon stacked local operations, may be suboptimal for this task. The process of downsampling and repeated local feature extraction in a deep CNN can progressively weaken or even lose these critical, non-local physical cues. Consequently, architectures with an innate global receptive field, like vision Transformers (ViTs)\cite{khan2022transformers}, appear to be a more natural fit. They are theoretically well-equipped to capture these dependencies from the outset. However, their global self-attention mechanism comes at the cost of quadratic complexity with respect to the input size, making them too computationally intensive for the real-time, high-resolution RM construction required by 6G systems. Moreover, ViTs often require extensive pre-training on massive datasets to develop effective visual representations\cite{dosovitskiy2020image}. This creates a critical dilemma: the architectural need for global context to respect the underlying physics is fundamentally at odds with the practical need for low-latency inference and precise local detail.

Structured state space models (SSMs) \cite{gu2021efficiently}, and particularly the Mamba architecture \cite{gu2023mamba}, have recently emerged as a promising approach to this problem. Mamba is specifically designed to model long-range dependencies with linear-time complexity, emulating the global context awareness of Transformers without the associated computational burden. The power of this approach has led to its swift adaptation in computer vision, with various works developing hybrid Mamba-CNN backbones for image-based tasks \cite{liu2024vision, hatamizadeh2025mambavision}. Furthermore, the ongoing development of even more efficient variants optimized for training speed and edge deployment underscores its practicality for real-world systems \cite{dao2024Transformers, he2025mobilemamba}. This combination of capabilities, achieving the global reach of Transformers with the efficiency of CNNs, suggests Mamba as a highly suitable architectural foundation for building models that are both highly accurate and fast enough for real-time RM construction.

Based on these capabilities, we propose \textbf{RadioMamba}, a hybrid Mamba-UNet architecture designed to resolve the accuracy-efficiency trade-off. The ability of Mamba to model global context with linear complexity offers a potential solution to the limitations of previous models. We hypothesize that by augmenting a computationally efficient, CNN-based architecture with the ability to model the long-range spatial dependencies that govern the underlying physics of radio propagation, it is possible to break the established trade-off, creating a model that can achieve state-of-the-art accuracy while maintaining the high computational efficiency required for real-time systems. RadioMamba is designed to test this hypothesis by embedding a hybrid Mamba-Convolutional block within the U-Net framework, where the Mamba branch models global interactions and a parallel convolutional branch extracts local features. The result is a single-pass model intended to improve upon prior methods in both performance and speed. Our primary contributions are as follows.

\begin{enumerate}
    \item We analyze the accuracy-efficiency trade-off in radio map construction, tracing it to the architectural limitations of current models. We identify that CNNs' local receptive fields fail to capture global wave physics, while global-context models like Transformers are computationally prohibitive for real-time use. Our analysis further establishes the unique suitability of the Mamba architecture, which models long-range dependencies with linear complexity, as a foundation to resolve this trade-off.
    \item We propose RadioMamba, a novel hybrid architecture that integrates the strengths of Mamba and CNNs within a U-Net framework. The core of our design is a synergistic Mamba-Convolutional block that captures both global long-range dependencies and fine-grained local features, making it uniquely suited for the physics-informed task of radio map construction.
    \item Our experiments show that RadioMamba establishes a new SOTA, outperforming existing methods like RadioDiff across all accuracy and perceptual metrics. It achieves this while being nearly 20 times faster and using only 2.9\% of the model parameters, demonstrating a resolution to the accuracy-efficiency trade-off and confirming its viability for real-time 6G systems.
\end{enumerate}

\begin{table}[t]
\centering
\captionsetup{font=small}
\caption{Comparison with Traditional and AI-enabled Paradigms.}
\label{tab:paradigm_comparison_final}
\resizebox{\columnwidth}{!}{%
    \renewcommand{\arraystretch}{1.3} 
    \begin{tabular}{@{}ccccc@{}}
        \toprule
        \textbf{Model/Method} & \textbf{AI-enabled} & \textbf{Global Context} & \textbf{Real-time Speed} & \textbf{High Accuracy} \\
        \midrule
        \cite{mao2018constructing} & \ding{55} & \ding{52} & \ding{55} & \ding{55} \\
        \cite{salski2020electromagnetic} & \ding{55} & \ding{52} & \ding{55} & \ding{52} \\
        \midrule
        \cite{levie2021radiounet} & \ding{52} & \ding{55} & \ding{52} & \ding{55} \\
        \cite{zhang2023rme} & \ding{52} & \ding{55} & \ding{52} & \ding{55} \\
        \cite{wang2024radiodiff} & \ding{52} & \ding{52} & \ding{55} & \ding{52} \\
        \textbf{Ours} & \ourmodelval{\ding{52}} & \ourmodelval{\ding{52}} & \ourmodelval{\ding{52}} & \ourmodelval{\ding{52}} \\
        \bottomrule
    \end{tabular}%
}
\end{table}

The remainder of this paper is organized as follows. Section \ref{sec:related_work} reviews related works. Section \ref{sec:system_model} presents the system model and problem formulation. Section \ref{sec:methodology} focus on the proposed RadioMamba architecture. Section \ref{sec:experiments} details the comprehensive experimental setup, results, and analyses. Finally, Section \ref{sec:conclusion} concludes the paper and discusses future research directions.

\section{Related Work}
\label{sec:related_work}
The methodologies for constructing RMs are broadly divided into two categories: measurement-based, which relies on physical data collection and interpolation \cite{mao2018constructing, cover1967nearest}, and sampling-free, which generates maps directly from environmental data. Our work focuses on the latter, a field that has been revolutionized by DL due to its power in solving complex generative tasks \cite{levie2021radiounet, zhang2023rme, wang2024radiodiff}. To provide a thorough background, this section reviews the pertinent literature across several key domains. We first survey the evolution of radio propagation modeling paradigms, then discuss the specific DL architectures for RM construction. Subsequently, we explore the broader context of AI-enabled wireless network optimization and the challenges of model efficiency, before finally delving into the state space models that form the technical foundation of our proposed method.

\subsection{Radio Propagation Modeling Paradigms}
The prediction of radio wave propagation is a long-standing challenge in wireless communications. The approaches can be broadly categorized into model-based (or physics-based) and data-driven methods \cite{g}.

\textbf{Model-Based Approaches.} These methods rely on fundamental physical principles to predict signal strength. They can be further divided into empirical and deterministic models.
Empirical models, such as the Okumura-Hata and Lee models, are derived from extensive measurement campaigns and provide statistical characterizations of path loss as a function of distance, frequency, and antenna heights \cite{a}. While computationally simple and widely used for large-scale network planning, their accuracy is limited as they generalize environmental characteristics into broad categories and may not capture site-specific details effectively \cite{k}.

Deterministic models, most notably those based on ray tracing, offer a more precise alternative by simulating the paths of electromagnetic waves as they reflect, diffract, and scatter within a detailed 3D model of the environment \cite{i, j}. Ray-tracing techniques can provide highly accurate, site-specific predictions for both narrowband and wideband channel characteristics, making them a powerful tool for designing and optimizing complex radio networks, from indoor WLANs to large-scale urban deployments \cite{e, l}. However, the accuracy of ray tracing is critically dependent on the quality and detail of the environmental database, and the computational cost can be prohibitive, especially for large areas or when considering a high number of interactions \cite{j}.

\textbf{Data-Driven Approaches.} With the proliferation of sensor data and advancements in machine learning, data-driven modeling has emerged as a powerful paradigm \cite{c}. Instead of relying on first principles, these methods learn the complex, nonlinear mapping directly from input-output data. In the context of wireless communications, this involves learning the relationship between environmental features and channel characteristics from measured or simulated channel data \cite{b}. The key advantage of data-driven methods is their ability to model highly complex systems where analytical or deterministic models are infeasible. Our work falls into this category, leveraging the power of deep learning to learn the mapping from environmental layouts to complete radio maps.

\subsection{Deep Learning for Radio Map Construction}
The evolution of DL for RM construction has seen a progression of architectures, each with distinct advantages and drawbacks.

\textbf{CNN Approaches.} CNNs, particularly encoder-decoder architectures, were among the first architectures applied to RM construction. \textbf{RadioUNet} \cite{levie2021radiounet} was one of the first models to apply the U-Net architecture\cite{ronneberger2015u} for RM construction. By representing the building layout and transmitter location as input channels, RadioUNet learns a direct mapping to the final pathloss map. The skip connections in U-Net are particularly effective, allowing the decoder to combine deep, semantic features with shallow, high-resolution features, which is essential for preserving sharp details. The primary strength of RadioUNet is its computational efficiency. However, its core convolution operator has an inherently local receptive field, limiting its ability to capture the global context of the radio environment and capping its ultimate performance.

\textbf{GAN Approaches.} To generate more perceptually realistic RMs, researchers turned to GANs \cite{goodfellow2020generative}. \textbf{RME-GAN} \cite{zhang2023rme} is a prime example, utilizing a conditional GAN (cGAN) framework\cite{mirza2014conditional} where a generator network creates RMs and a discriminator distinguishes them from ground-truth maps. Other works have also explored GANs for fast and accurate cooperative radio map estimation \cite{zhang2024fast}. This adversarial process often results in maps with sharper details. GANs are also computationally efficient during inference, but their training can be unstable and requires careful tuning.

\textbf{Advanced Generative Models.} Recently, denoising diffusion probabilistic models (DDPMs) \cite{ho2020denoising} have established SOTA performance. \textbf{RadioDiff} \cite{wang2024radiodiff} is the first work to adapt this paradigm to RM construction. It formulates the task as a conditional generation process, starting from pure random noise and iteratively denoises it over hundreds of timesteps to produce a high-fidelity RM. This iterative refinement allows diffusion models to achieve high accuracy. However, this improved performance comes at the cost of high computational complexity and slow inference, making them orders of magnitude slower than single-pass models and posing challenges for real-time deployment. The broader field of wireless communications has also seen a surge in the application of generative AI (GAI), for instance, to construct channel knowledge maps (CKMs) \cite{fu2025ckmdiff} or to enhance physical layer security, demonstrating the versatility of these models beyond pure generation tasks \cite{4}.

\subsection{AI-Enabled Wireless Network Optimization}
The construction of an accurate RM is often not an end in itself, but a means to enable more intelligent network management and optimization. The application of AI, particularly reinforcement learning (RL) and its variants, to wireless communications is a burgeoning field.

In 6G networks, which are envisioned as a `network of subnetworks,' dynamic resource management becomes exceptionally challenging due to high mobility and mutual interference. Multi-agent reinforcement learning (MARL) has been proposed to address this, enabling subnetworks to learn cooperative policies for channel selection and power control. Advanced MARL architectures, such as those incorporating graph attention networks, can effectively reason about inter-subnetwork relationships using only accessible information like RSSI, avoiding the need for hard-to-obtain global channel state information \cite{h}. Similarly, in D2D-enabled 6G networks, federated reinforcement learning provides a framework for decentralized resource allocation that maximizes network capacity while respecting user privacy \cite{d}.

Accurate channel estimation is another critical task where AI is making inroads. In Massive MIMO systems, for instance, data-driven approaches using neural networks are being explored to estimate effective downlink channel gains without the need for downlink pilots, outperforming traditional model-based methods, especially in environments with low channel hardening \cite{g}. This trend of AI-enabled, data-driven channel modeling, which includes diverse applications from radio map construction to modulation classification \cite{7}, is seen as a key enabler for future communication systems, promising to deliver more accurate and adaptable models than classical methods \cite{b}.

\subsection{Model Efficiency and Edge Deployment}
A major bottleneck for deploying advanced AI models in real-world wireless systems, particularly on edge devices like UAVs and vehicles, is their computational and communication overhead. This has spurred significant research into model efficiency.

Federated Learning (FL), while designed to protect data privacy by training models locally on devices, introduces a significant communication burden due to the repeated transmission of model parameters between clients and a central server. To mitigate this, various model compression techniques have been investigated. These techniques aim to reduce the size of the models being transmitted, for both upstream (client-to-server) and downstream (server-to-client) communication, by creating sparse models from dense ones without significant loss in accuracy \cite{f}. Hierarchical FL architectures, which introduce intermediate aggregation servers, have also been proposed to reduce communication costs and better handle data heterogeneity in large-scale systems like the Industrial IoT \cite{10}.

The challenge of deploying AI extends to Digital Twins (DTs) as well, where tasks like adaptive caching are crucial for optimizing performance in heterogeneous IoT environments \cite{6}. The construction and maintenance of high-fidelity DTs for complex systems like air-ground networks or vehicular edge networks require substantial resources. Consequently, research has focused on intelligent task offloading and resource management. For instance, federated deep reinforcement learning has been used to optimize task offloading in DT edge networks \cite{9}, while incentive mechanisms based on game theory have been designed to encourage participation in FL for dynamic DTs in air-ground networks \cite{8}. These efforts, spanning from general IoT to specific domains like secure vehicular networks \cite{5}, highlight the critical need for communication- and computation-efficient AI solutions, which directly motivates our work on developing a lightweight yet powerful architecture for RM construction.

\subsection{State Space Models and Mamba}
Our work is directly inspired by recent advances in structured SSMs, which have recently emerged as a powerful and highly efficient alternative to Transformers for modeling long-range dependencies.

\subsubsection{Foundations of State Space Models}
SSMs\cite{gu2021efficiently}, originating from classical control theory, model a system by mapping a 1-D input sequence $\mathbf{u}(t)$ to an output $\mathbf{y}(t)$ via a latent state vector $\mathbf{h}(t) \in \mathbb{R}^{N}$. A continuous-time SSM is defined by the linear ordinary differential equations (ODEs) as follows.
\begin{align}
    \frac{d\mathbf{h}(t)}{dt} &= \mathbf{A}\mathbf{h}(t) + \mathbf{B}\mathbf{u}(t) \label{eq:ssm_h_cont_rebuttal}, \\
    \mathbf{y}(t) &= \mathbf{C}\mathbf{h}(t) + \mathbf{D}\mathbf{u}(t) \label{eq:ssm_y_cont_rebuttal},
\end{align}
where $\mathbf{A} \in \mathbb{R}^{N \times N}$ is the state matrix that governs the internal state dynamics, and $\mathbf{B} \in \mathbb{R}^{N \times 1}$, $\mathbf{C} \in \mathbb{R}^{1 \times N}$ are projection matrices that map the input to the state and the state to the output, respectively.

For use in deep learning with discrete data, such as tokens or pixels, these continuous parameters must be discretized. A fixed sampling timestep $\Delta$ is introduced, and a discretization rule, such as the zero-order hold (ZOH), transforms the continuous parameters into their discrete counterparts, $\bar{\mathbf{A}}$ and $\bar{\mathbf{B}}$, which is as follows.
\begin{align}
    \bar{\mathbf{A}} &= \exp(\Delta \mathbf{A}), \\
    \bar{\mathbf{B}} &= (\Delta \mathbf{A})^{-1}(\exp(\Delta \mathbf{A}) - \mathbf{I})\Delta \mathbf{B},
\end{align}
The SSM can then be computed either as a standard recurrence, which is efficient for inference, or a global convolution, which is efficient for training. It can be formulated as follows.
\begin{align}
    \mathbf{h}_k &= \bar{\mathbf{A}}\mathbf{h}_{k-1} + \bar{\mathbf{B}}\mathbf{u}_k \label{eq:ssm_h_disc_rebuttal}, \\
    \mathbf{y}_k &= \mathbf{C}\mathbf{h}_k \label{eq:ssm_y_disc_rebuttal}, \\
    \bar{\mathbf{K}}_k &= \mathbf{C}\bar{\mathbf{A}}^k\bar{\mathbf{B}}, \\
    \mathbf{y} &= \mathbf{u} * \bar{\mathbf{K}} \label{eq:ssm_conv_rebuttal}.
\end{align}
While powerful, the computational cost of these formulations, especially for long sequences, limited early SSMs. The breakthrough came with structured variants, like S4 \cite{gu2021efficiently}, which imposed structure (e.g., diagonal plus low-rank) on the $\mathbf{A}$ matrix. This allowed for the efficient computation of the convolutional kernel $\bar{\mathbf{K}}$ in near-linear time, enabling SSMs to rival Transformers on long-sequence benchmarks.

\subsubsection{Mamba: Selective State Spaces and Efficiency}
\textbf{Mamba} \cite{gu2023mamba} represents a advancement in this area, overcoming a critical limitation of prior SSMs: their time- and input-invariance. Models like S4 use fixed parameters that are not data-dependent, meaning they cannot adapt their behavior to the specific input content. Mamba addresses this with two key innovations:

\paragraph{Selection Mechanism}
The key advantage of Mamba is its ability to perform content-based reasoning by making the SSM parameters input-dependent. For each token $\mathbf{u}_k$ in the input sequence, Mamba dynamically generates specific parameters $\Delta_k$, $\mathbf{B}_k$, and $\mathbf{C}_k$ via linear projections from the input as follows.
\begin{align}
    \Delta_k &= \text{softplus}(\text{Linear}_{\Delta}(\mathbf{u}_k)), \\
    \mathbf{B}_k &= \text{Linear}_B(\mathbf{u}_k) \in \mathbb{R}^{N \times 1}, \\
    \mathbf{C}_k &= \text{Linear}_C(\mathbf{u}_k) \in \mathbb{R}^{1 \times N},
\end{align}
where $\text{Linear}_{\Delta}$, $\text{Linear}_B$, and $\text{Linear}_C$ are learnable projection layers. Note that the core state matrix $\mathbf{A} \in \mathbb{R}^{N \times N}$ remains fixed. These dynamic parameters are then used to compute time-varying discrete matrices $\bar{\mathbf{A}}_k$ and $\bar{\mathbf{B}}_k$ for each step as follows.
\begin{align}
    \bar{\mathbf{A}}_k &= \exp(\Delta_k \mathbf{A}), \\
    \bar{\mathbf{B}}_k &= (\Delta_k \mathbf{A})^{-1}(\exp(\Delta_k \mathbf{A}) - \mathbf{I})\Delta_k \mathbf{B}_k,
\end{align}
This results in a selective SSM recurrence where the state update is modulated by the input at every step as follows.
\begin{align}
    \mathbf{h}_k &= \bar{\mathbf{A}}_k \mathbf{h}_{k-1} + \bar{\mathbf{B}}_k \mathbf{u}_k, \\
    \mathbf{y}_k &= \mathbf{C}_k \mathbf{h}_k.
\end{align}
This dynamic parameterization allows the model to selectively propagate or forget information in its latent state $\mathbf{h}_k$. By learning to modulate $\Delta_k$, for instance, the model can decide whether to focus on fine-grained local information (small $\Delta_k$) or long-range dependencies (large $\Delta_k$). This selectivity gives Mamba the context-aware reasoning of attention mechanisms without the quadratic computational cost.

\paragraph{Hardware-Aware Parallel Algorithm}
While the selection mechanism makes the model powerful, it breaks the time-invariance that allowed for efficient convolutional computation (Eq. \ref{eq:ssm_conv_rebuttal}). A naive recurrent implementation (Eq. \ref{eq:ssm_h_disc_rebuttal}) would be slow to train on parallel hardware like GPUs. Mamba's second core innovation is a \textbf{hardware-aware algorithm} that leverages the associative property of its scan operation. It expresses the computation in a way that can be parallelized for training, similar to prefix sums, while still collapsing back to a highly efficient recurrent form for autoregressive inference. This dual formulation allows Mamba to be trained at high speed on modern accelerators while retaining linear-time inference.

These innovations combined give Mamba the ability to model extremely long-range dependencies with a computational complexity that is linear in sequence length. This characteristic is particularly suitable for radio map construction. The radio environment is a system where the `state' (pathloss strength) at one location is dependent on a long `sequence' of interactions with obstacles across the entire map. Mamba's ability to efficiently process this global context makes it uniquely suited for this task. While originally designed for 1D sequences, recent works like Vision Mamba \cite{liu2024vmamba, liu2024vision} have adapted it to 2D vision tasks by scanning images along multiple directions. Our work is inspired by this principle, adapting Mamba into a lightweight and synergistic architecture tailored for the physics-based generation task of radio map construction.

\begin{figure*}[t]
    \centering
    \captionsetup{font=small}
    \includegraphics[width=0.95\textwidth]{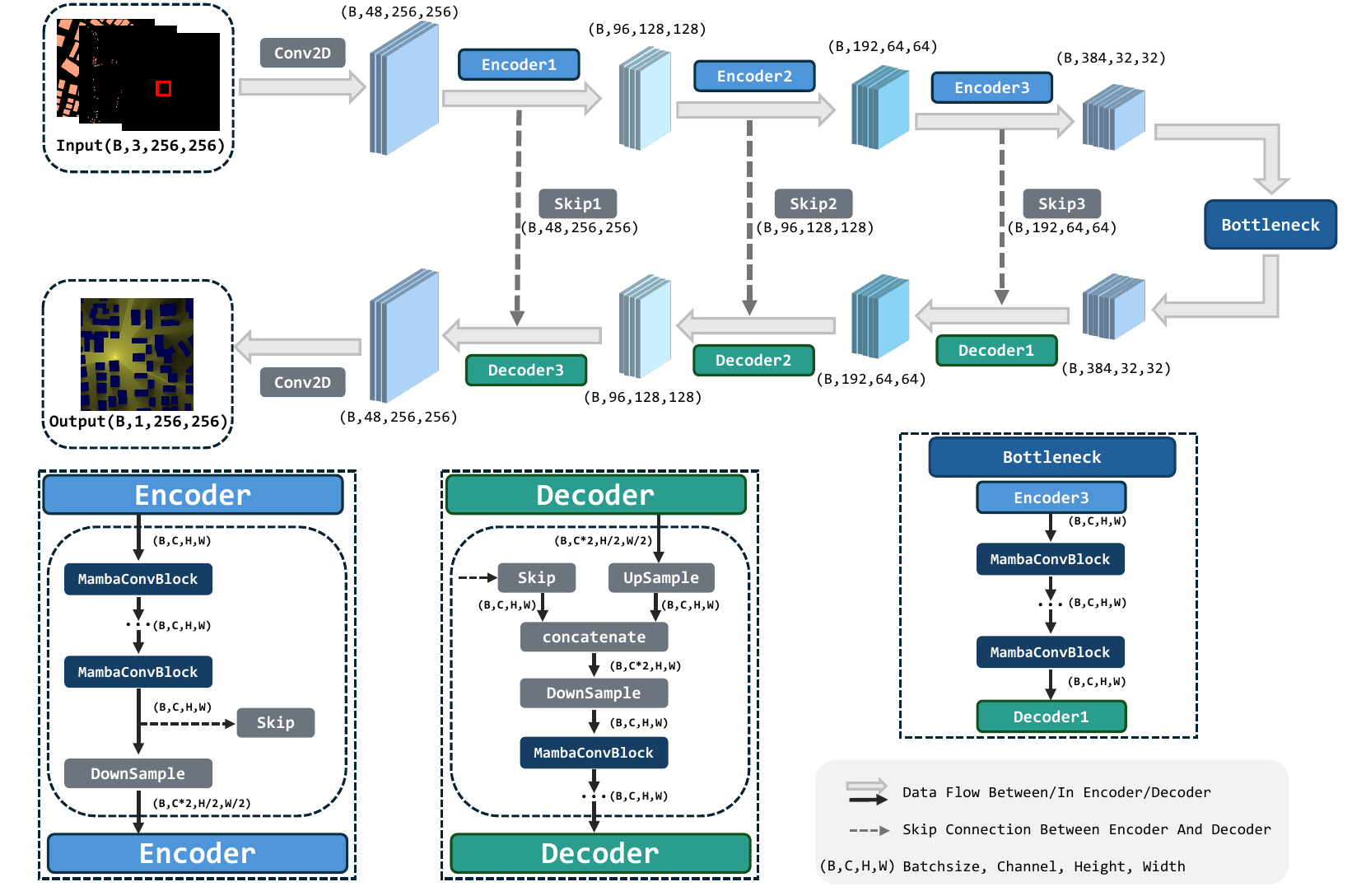}
    \caption{The overall architecture of RadioMamba.}
    \label{fig:architecture}
    \vspace{-12pt}
\end{figure*}

\begin{figure}[ht]
    \centering
    \captionsetup{font=small}
    \includegraphics[width=0.98\columnwidth]{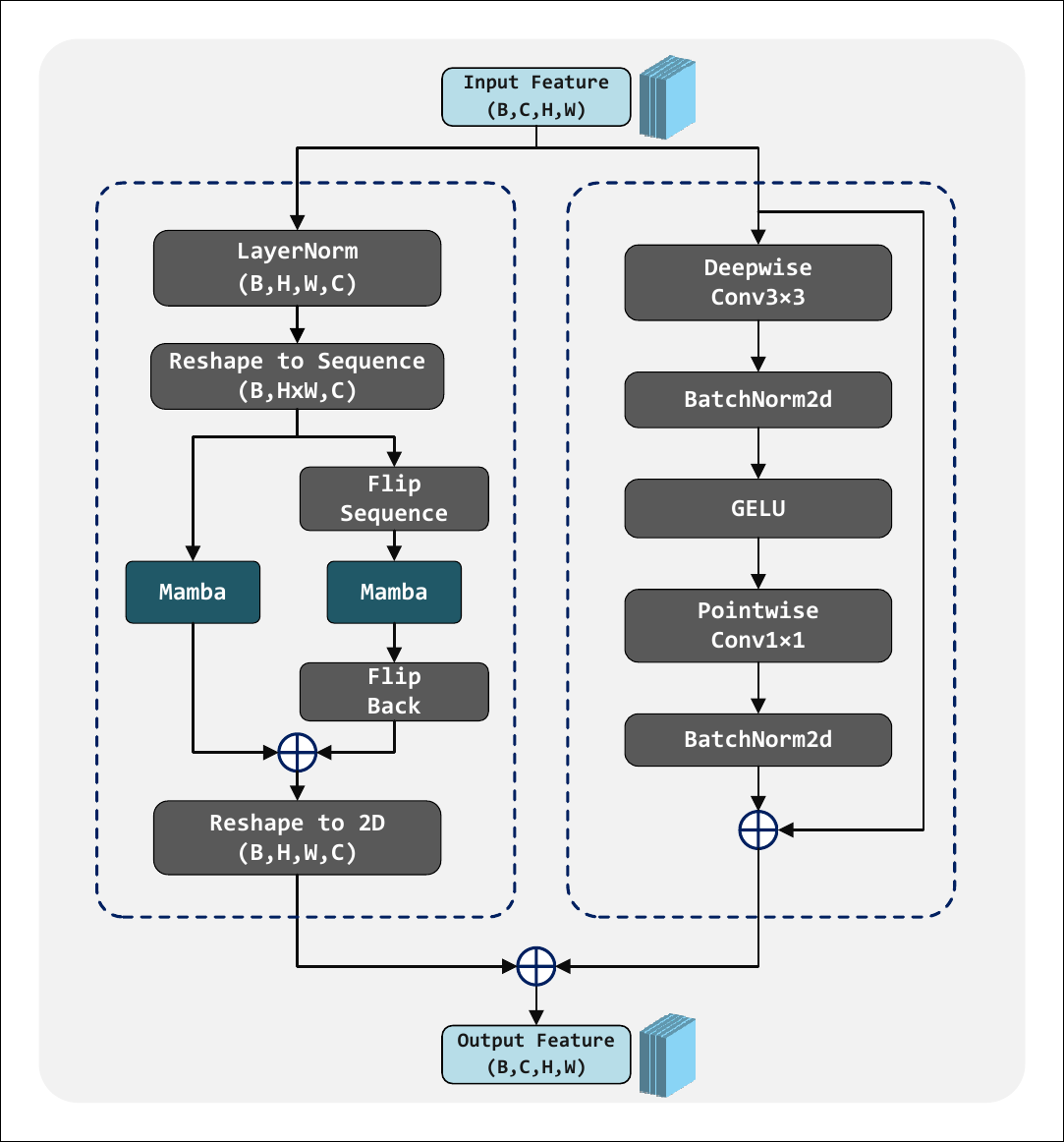}
    \caption{Schematic of the proposed MambaConvBlock.}
    \label{fig:mambaconvblock}
    \vspace{-12pt}
\end{figure}


\section{System Model and Problem Formulation}
\label{sec:system_model}

In this section, we formally define the environment for sampling-free radio map construction and formulate the task as a conditional image generation problem, drawing upon the detailed setup from prior work such as RadioDiff \cite{wang2024radiodiff} for consistency and comparability.

\subsection{Environment and Radio Map Representation}
We consider a geographical area discretized into a 2D grid of size $N \times N$, where each grid cell (pixel) represents a small square area (e.g., $1 \text{m} \times 1 \text{m}$). The environment within this grid is characterized by the presence of various obstacles that affect radio wave propagation.

\textbf{Input Features}: The model is conditioned on the physical layout of the environment, which is provided as a multi-channel tensor. This input, denoted as $\mathbf{C}_{in} \in \mathbb{R}^{C \times N \times N}$, is composed of several channels that describe the scene:
\begin{itemize}
    \item \textbf{Static Obstacle Map ($\mathbf{H}_s$)}: A binary matrix of size $N \times N$. An element $h_{i,j}^s = 1$ indicates the presence of a permanent, static obstacle like a building at location $(i,j)$, while $h_{i,j}^s = 0$ signifies free space. This map provides the primary, unchanging geometry of the environment.
    \item \textbf{Dynamic Obstacle Map ($\mathbf{H}_d$)}: A binary matrix of size $N \times N$ that captures transient objects, such as vehicles. An element $h_{i,j}^d = 1$ indicates the presence of a dynamic obstacle. This channel allows the model to account for temporary changes in the environment, which is crucial for dynamic RM (DRM) construction. For static RM (SRM) construction, this channel may be zero-filled or omitted.
    \item \textbf{Transmitter Location Map ($\mathbf{R}$)}: A binary matrix of size $N \times N$ that serves as a one-hot encoding of the transmitter's position. The pixel corresponding to the transmitter's coordinates $(x_{tx}, y_{tx})$ has a value of 1, and all other pixels are 0.
\end{itemize}
These maps, representing different aspects of the physical world, are concatenated channel-wise to form the complete input tensor for the neural network. In our experiments, following the RadioMapSeer benchmark \cite{levie2021radiounet}, $N=256$.

\textbf{Output Radio Map ($\mathbf{P}$)}: The model's objective is to generate the radio map, which is a single-channel, real-valued matrix $\mathbf{P} \in \mathbb{R}^{N \times N}$. Each element $p_{i,j}$ in this matrix represents the predicted pathloss value in decibels (dB) at the grid location $(i, j)$. This dense map provides a comprehensive characterization of the signal coverage throughout the area for a given transmitter location and environmental state. Pathloss is typically a large negative value, which is normalized to a grayscale range (e.g., [0, 1]) for model training and prediction.

\subsection{Problem Formulation}
The task of radio map construction is to learn a deterministic function, $\mathcal{F}_{\theta}$, parameterized by a set of learnable weights $\theta$, that maps the multi-channel environmental input $\mathbf{C}_{\text{in}}$ to its corresponding radio map $\mathbf{P}$. This function must implicitly learn the complex underlying physics of radio wave propagation, including reflection, diffraction, and scattering. The problem can be formally stated as:

\noindent\textbf{Problem 1. (Radio Map construction)}
Given a dataset $\mathcal{D} = \{(\mathbf{C}_{\text{in}}^{(i)}, \mathbf{P}^{(i)})\}_{i=1}^{M}$ consisting of $M$ pairs of environmental configuration tensors and their corresponding ground-truth radio maps. The goal is to find the optimal set of parameters $\theta^*$ for a deep neural network $\mathcal{F}_{\theta}$ that minimizes a predefined loss function $\mathcal{L}(\cdot, \cdot)$ over the dataset:
\begin{equation}
    \theta^* = \arg\min_{\theta} \mathbb{E}_{(\mathbf{C}_{\text{in}}, \mathbf{P}) \sim \mathcal{D}} \left[ \mathcal{L}(\mathcal{F}_{\theta}(\mathbf{C}_{\text{in}}), \mathbf{P}) \right].
\end{equation}
where $\hat{\mathbf{P}} = \mathcal{F}_{\theta}(\mathbf{C}_{\text{in}})$ is the predicted radio map.

This task is fundamentally a \textbf{complex spatial mapping problem}. The network's objective is to learn a deterministic function that translates an input (a single transmitter point and a building layout) into a complete, continuous-valued spatial distribution, the radio map $\hat{\mathbf{P}}$. The pathloss value at any given location $(i,j)$ is not merely a function of its local neighborhood; it is dependent on the \textbf{entire global geometry} of the environment. Obstacles hundreds of meters away can cast long shadows (non-line-of-sight conditions) or create complex multi-path reflections that fundamentally alter the signal strength. This non-local nature of wave physics requires that a successful model must possess a powerful mechanism for capturing long-range spatial dependencies. A model with only a limited, local receptive field is inherently limited, as it cannot `see' the distant obstacles that are crucial for an accurate prediction. This physical requirement motivated our design of RadioMamba.


\section{The Proposed RadioMamba}
\label{sec:methodology}
To resolve the performance-efficiency dilemma, we designed RadioMamba, an architecture that leverages the hierarchical feature extraction of U-Net while enhancing its feature representation capabilities with a hybrid block that efficiently models both global and local spatial features.

\subsection{Overall Architecture}
The overall architecture of RadioMamba is based on the well-established encoder-decoder design of U-Net, illustrated in Fig. \ref{fig:architecture}. This structure is inherently suited for image-to-image tasks due to its ability to preserve spatial information across multiple scales.
\begin{itemize}
    \item \textbf{Input}: The model accepts a multi-channel input tensor of size $256 \times 256$, representing the building map, transmitter location, and an auxiliary channel for inputs such as dynamic obstacles. An initial convolution projects this input into a high-dimensional feature space.
    \item \textbf{Encoder}: The encoder path consists of three stages. Each stage comprises a series of our custom \textit{MambaConvBlocks} followed by a downsampling operation using a $2 \times 2$ strided convolution that halves the spatial resolution (e.g., $256 \to 128 \to \dots \to 32$) while doubling the number of feature channels. This process builds a feature representation that captures increasingly abstract semantic information.
    \item \textbf{Bottleneck}: At the lowest resolution ($32 \times 32$), the feature map passes through the final set of MambaConvBlocks, acting as the bottleneck which processes the most compressed and semantically rich feature representation.
    \item \textbf{Decoder}: The decoder path symmetrically mirrors the encoder. It consists of three stages that use transposed convolutions to progressively upsample the feature maps back to the original resolution. At each stage, the upsampled feature map is concatenated with the corresponding high-resolution feature map from the encoder via a skip connection. A $1 \times 1$ convolution first fuses the concatenated features, which are then processed by a series of MambaConvBlocks to refine the representation.
    \item \textbf{Output}: A final $1 \times 1$ convolutional layer projects the feature map from the last decoder stage back to the desired single output channel for the radio map, producing the final prediction.
\end{itemize}
The core innovation of our model lies not in the U-Net skeleton itself, but in the effective building blocks used at each stage of the encoder and decoder.

\subsection{The MambaConvBlock}
Standard U-Nets use blocks of convolutional layers. To overcome their limited receptive field, we designed the \textbf{MambaConvBlock}, a hybrid module that operates with two parallel branches to process information concurrently, as shown in Fig. \ref{fig:mambaconvblock}. This design allows the block to benefit from the best of both worlds: the local inductive bias of CNNs and the global receptive field of SSMs.

\subsubsection{Local Feature Branch}
This branch is engineered to extract local textures and high-frequency details, which are important for delineating the sharp, physically-correct shadow boundaries around environmental obstacles. To achieve this with maximum computational efficiency, the branch is constructed as a ResidualConvBlock which leverages the principle of depthwise separable convolutions. This modern architectural pattern factorizes a standard convolution into two distinct, more efficient operations.

Let the input tensor be $\mathbf{X} \in \mathbb{R}^{C_{in} \times H \times W}$. A standard 2D convolution would directly map this to an output with a computational cost, measured in multiply-accumulate operations (MACs) approximately as follows.
\begin{equation}
    \text{Cost}_{\text{std}} = H \cdot W \cdot C_{in} \cdot C_{out} \cdot K_h \cdot K_w,
\end{equation}
Depthwise separable convolutions reduce this cost by splitting the process.

\paragraph{Depthwise Convolution}
The first step is a depthwise convolution, which applies a single spatial filter to each input channel independently. This operation handles the spatial filtering, capturing patterns like edges, but does not mix information across channels. Its cost is as follows.
\begin{equation}
    \text{Cost}_{\text{dw}} = H \cdot W \cdot C_{in} \cdot K_h \cdot K_w,
\end{equation}

\paragraph{Pointwise Convolution}
The second step is a pointwise convolution (a $1\times1$ convolution), which projects the intermediate features from the depthwise step into the final output channel space by linearly combining features across channels. This step is responsible for feature mixing, and its cost is as follows.
\begin{equation}
    \text{Cost}_{\text{pw}} = H \cdot W \cdot C_{in} \cdot C_{out},
\end{equation}

The primary motivation for this factorization is the reduction in computational load. The ratio of the separable cost ($\text{Cost}_{\text{dw}} + \text{Cost}_{\text{pw}}$) to the standard cost is given as follows.
\begin{equation}
    \frac{\text{Cost}_{\text{dw}} + \text{Cost}_{\text{pw}}}{\text{Cost}_{\text{std}}} = \frac{1}{C_{out}} + \frac{1}{K_h K_w},
\end{equation}

For a typical $3\times3$ kernel and a large number of output channels, this ratio approaches $\frac{1}{9}$, implying an almost 9-fold reduction in MACs and parameters.

The entire operation is wrapped in a residual connection. Let $\mathcal{F}_{conv}$ represent the sequence of depthwise convolution, a GELU activation, and a pointwise convolution. The output of the ResidualConvBlock $\mathbf{Y}_{conv}$ is as follows.
\begin{equation}
    \mathbf{Y}_{conv} = \mathbf{X} + \mathcal{F}_{conv}(\mathbf{X}).
\end{equation}

This residual connection is crucial for enabling stable training in a deep architecture, as it facilitates unimpeded gradient flow and ensures that fine-grained features are preserved and progressively refined.

\begin{table*}[ht]
    \centering
    \captionsetup{font=small}
    \caption{Quantitative Comparison for SRM and DRM construction.}
    \label{tab:combined_rm_results}
    \resizebox{1.4\columnwidth}{!}{%
        \renewcommand{\arraystretch}{1.4} 
        \begin{tabular}{c|lcccc}
            \toprule
            \multicolumn{2}{c}{\textbf{Model}} & \textbf{NMSE} $\downarrow$ & \textbf{RMSE (dB)} $\downarrow$ & \textbf{SSIM} $\uparrow$ & \textbf{PSNR (dB)} $\uparrow$ \\
            \midrule
            \multirow{4}{*}{\textbf{SRM}} 
            & RME-GAN \cite{zhang2023rme} & 0.0096 & 0.0279 & 0.9431 & 31.35 \\
            & RadioUNet \cite{levie2021radiounet} & 0.0088 & 0.0266 & 0.9466 & 31.77 \\
            & RadioDiff \cite{wang2024radiodiff} & \secondbestval{0.0072} & \secondbestval{0.0240} & \secondbestval{0.9560} & \secondbestval{32.67} \\
            & \textbf{RadioMamba(Ours)} & \ourmodelval{0.0050} & \ourmodelval{0.0199} & \ourmodelval{0.9673} & \ourmodelval{34.32} \\
            \cmidrule(lr){1-6} 
            \multirow{4}{*}{\textbf{DRM}} 
            & RME-GAN & 0.0115 & 0.0306 & 0.9276 & 30.42 \\
            & RadioUNet & 0.0107 & 0.0291 & 0.9291 & 30.89 \\
            & RadioDiff & \secondbestval{0.0090} & \secondbestval{0.0266} & \secondbestval{0.9432} & \secondbestval{31.71} \\
            & \textbf{RadioMamba(Ours)} & \ourmodelval{0.0063} & \ourmodelval{0.0221} & \ourmodelval{0.9580} & \ourmodelval{33.33} \\
            \bottomrule
        \end{tabular}%
    }
    \vspace{-6pt}
\end{table*}

\begin{table}[ht]
    \centering
    \captionsetup{font=small}
    \caption{Efficiency and Complexity Comparison.}
    \label{tab:efficiency}
    \resizebox{1.0\columnwidth}{!}{%
        \renewcommand{\arraystretch}{1.5} 
        \begin{tabular}{lccc}
            \toprule
            \textbf{Model} & \textbf{Time (s)} $\downarrow$ & \textbf{Params (M)} $\downarrow$ & \textbf{Memory (MB)} $\downarrow$ \\
            \midrule
            RME-GAN & 0.0045 & 16.04 & 527 \\
            RadioUNet & 0.0041 & 13.27 & 525 \\
            RadioDiff & 0.5535 & 297.74 & 2067 \\
            \textbf{RadioMamba(Ours)} & \ourmodelval{0.0280} & \ourmodelval{8.60} & \ourmodelval{808} \\
            \cmidrule(lr){2-4}
             & \textbf{\textcolor{darkgreen}{$\downarrow$94.94\%}} & \textbf{\textcolor{darkgreen}{$\downarrow$97.11\%}} & \textbf{\textcolor{darkgreen}{$\downarrow$60.91\%}} \\
            \bottomrule
        \end{tabular}%
    }
    \vspace{-12pt}
\end{table}

\subsubsection{Global Context Branch}
This branch is the key to capturing long-range dependencies across the entire feature map, modeling the global nature of wave propagation. It employs our Mamba module, which we term \textbf{SS2D-Mamba}. To adapt the inherently 1D Mamba model to 2D image data, we first transform the input feature map into a sequence and then perform a bidirectional scan for comprehensive context aggregation\cite{liu2024vmamba}.

Given an input feature tensor $\mathbf{X} \in \mathbb{R}^{B \times C \times H \times W}$, it is first normalized using LayerNorm for stable processing. It is then unrolled into a sequence of length $L = H \times W$ via a standard raster-scan flattening operation, which we denote as $\mathcal{S}$. This yields the sequence $\mathbf{x}_{\text{seq}} \in \mathbb{R}^{B \times L \times C}$ as follows.
\begin{equation}
    \mathbf{x}_{\text{seq}} = \mathcal{S}(\text{LayerNorm}(\mathbf{X})),
\end{equation}

This sequence is then processed by the core Mamba model, which we represent with the operator $\mathcal{M}$, in both forward and backward directions to ensure that the context for each token is non-causal and includes information from the entire sequence. Let $\mathcal{R}$ define the sequence reversal operator, which flips a sequence along its length dimension. The forward and backward outputs are then computed as follows.
\begin{align}
    \mathbf{y}_{\text{fwd}} &= \mathcal{M}(\mathbf{x}_{\text{seq}}) ,\\
    \mathbf{y}_{\text{bwd}} &= \mathcal{R}\left( \mathcal{M}\left( \mathcal{R}(\mathbf{x}_{\text{seq}}) \right) \right).
\end{align}
The outputs from these two directional scans are then aggregated, typically via element-wise addition, and the resulting sequence is reshaped back into a 2D feature map $\mathbf{X}_{\text{mamba}} \in \mathbb{R}^{B \times C \times H \times W}$. This process allows every pixel in the feature map to efficiently integrate information from all other pixels, providing a comprehensive global context.

We acknowledge a theoretical limitation in this approach: radio wave propagation is physically isotropic, meaning it spreads uniformly in all directions, whereas the raster-scan flattening method is anisotropic, prioritizing horizontal and vertical spatial relationships. However, we hypothesize that this limitation is effectively mitigated by several factors in our architecture. First, the bidirectional scan captures dependencies in two primary directions. Second, the hierarchical structure of the U-Net, which processes features at multiple scales, helps to integrate contextual information more holistically. Finally, the parallel convolutional branch, with its strong local and isotropic inductive bias, complements the Mamba branch by reinforcing spatially-agnostic local features. Future work could explore more advanced, isotropic scanning patterns, such as space-filling curves, to potentially align the model even more closely with the underlying physics.

\subsubsection{Branch Fusion}
The outputs from the two parallel branches are synergistically combined via simple element-wise addition. This operation fuses the global, long-range context captured by the Mamba branch with the local, high-frequency details extracted by the convolutional branch. The outputs are as follows.
\begin{equation}
    \mathbf{Y}_{out} = \mathbf{Y}_{conv} + \mathbf{X}_{mamba}.
\end{equation}
This MambaConvBlock serves as the fundamental building unit throughout our RadioMamba architecture, enabling it to achieve better feature representation at every level of the hierarchy.

\section{Experiments}
\label{sec:experiments}

\begin{figure*}[ht]
    \centering
    \captionsetup{font=small}
    \begin{tabular}{c}
        \begin{adjustbox}{valign=t}
            \begin{tabular}{ccccc}
                \includegraphics[width=0.18\linewidth]{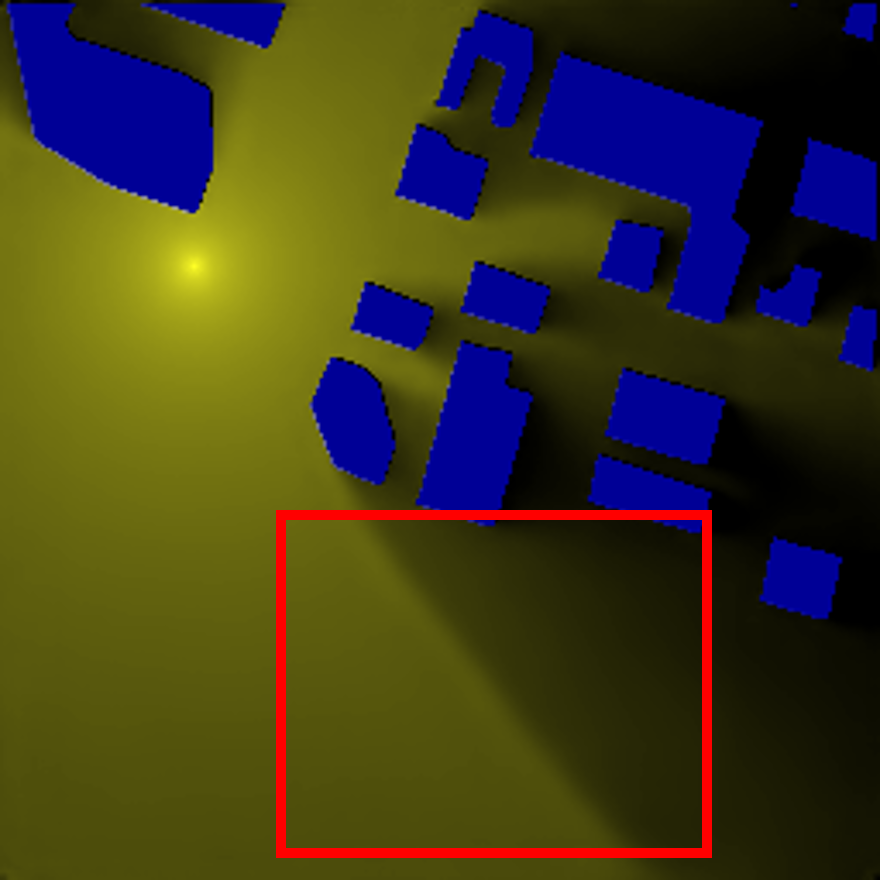}   \hspace{-4mm} &
                \includegraphics[width=0.18\linewidth]{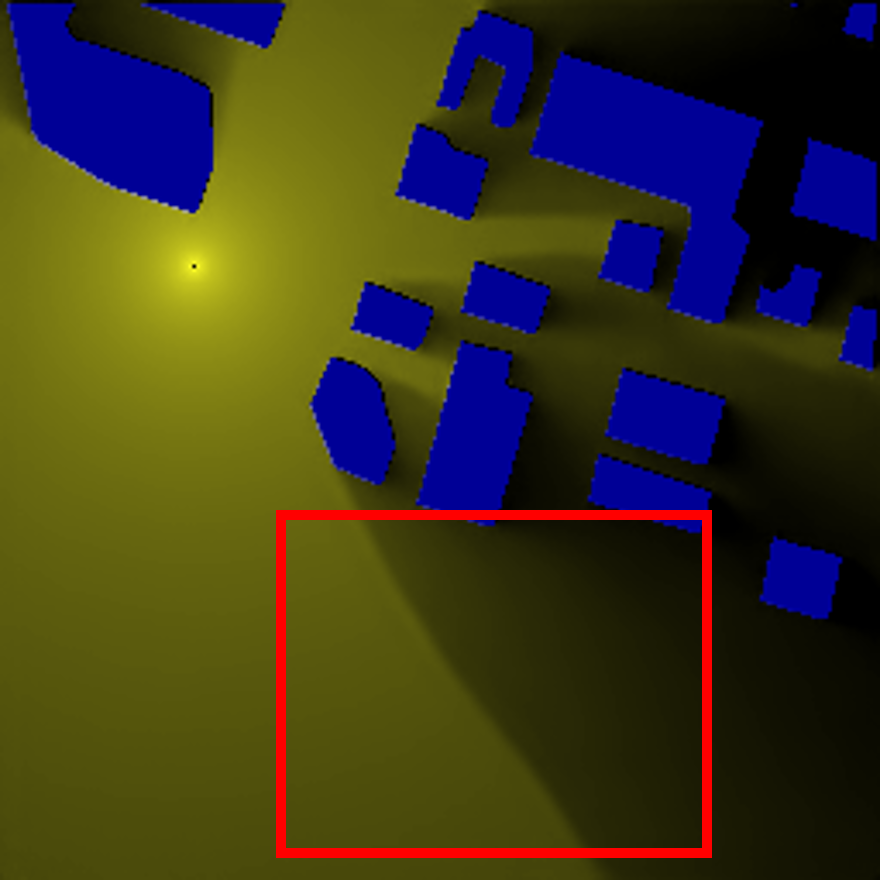}  \hspace{-4mm} &
                \includegraphics[width=0.18\linewidth]{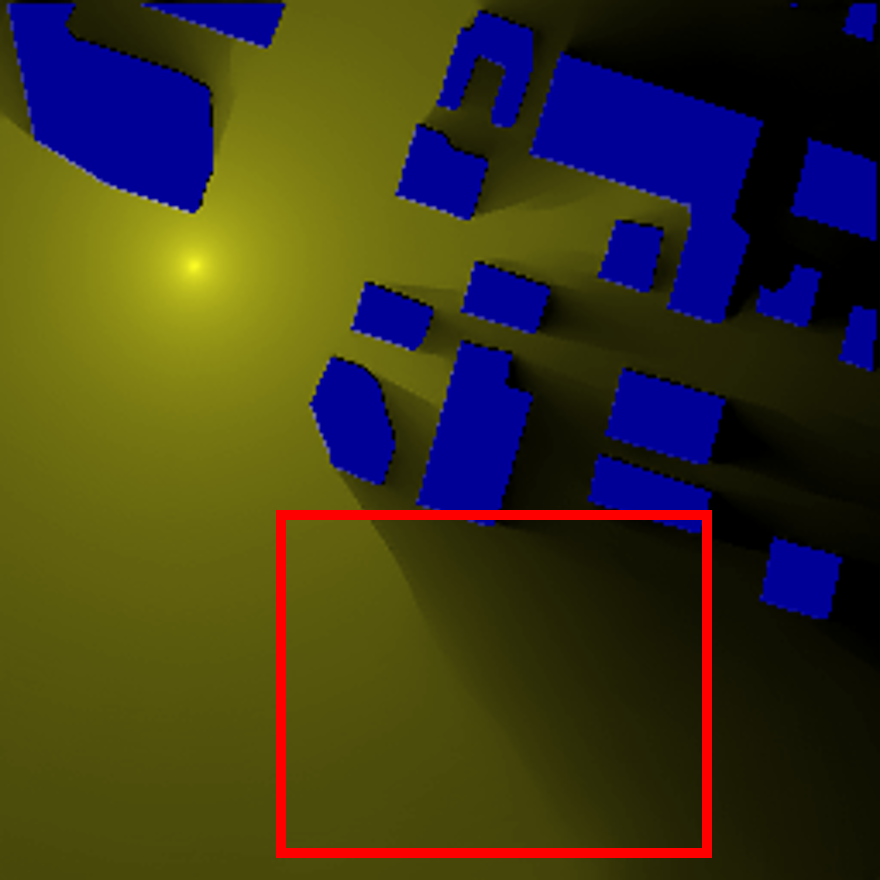}  \hspace{-4mm} &
                \includegraphics[width=0.18\linewidth]{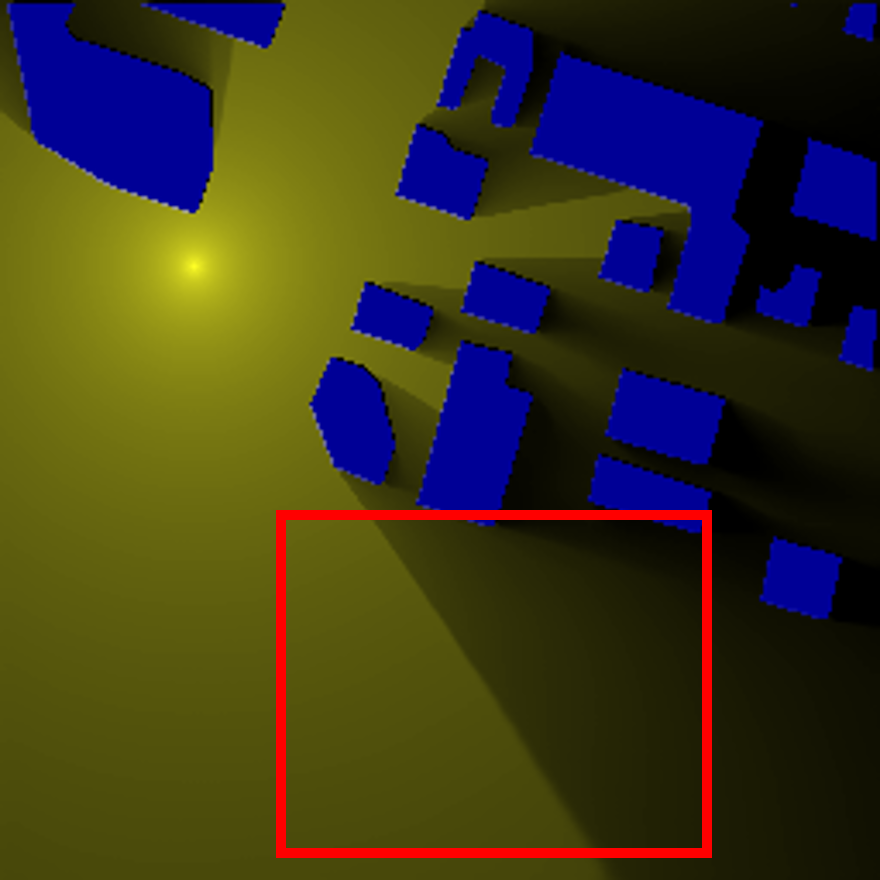} \hspace{-4mm} &
                \includegraphics[width=0.18\linewidth]{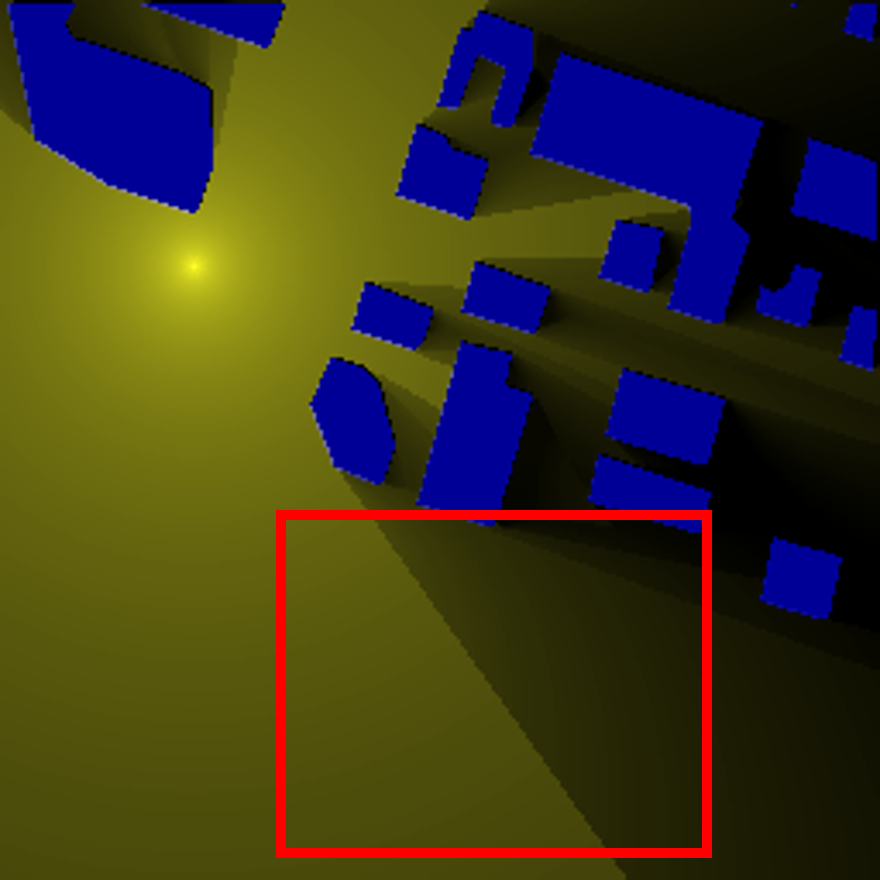}
            \end{tabular}
        \end{adjustbox} \\
        \vspace{1mm} 

        \begin{adjustbox}{valign=t}
            \begin{tabular}{ccccc}
                \includegraphics[width=0.18\linewidth]{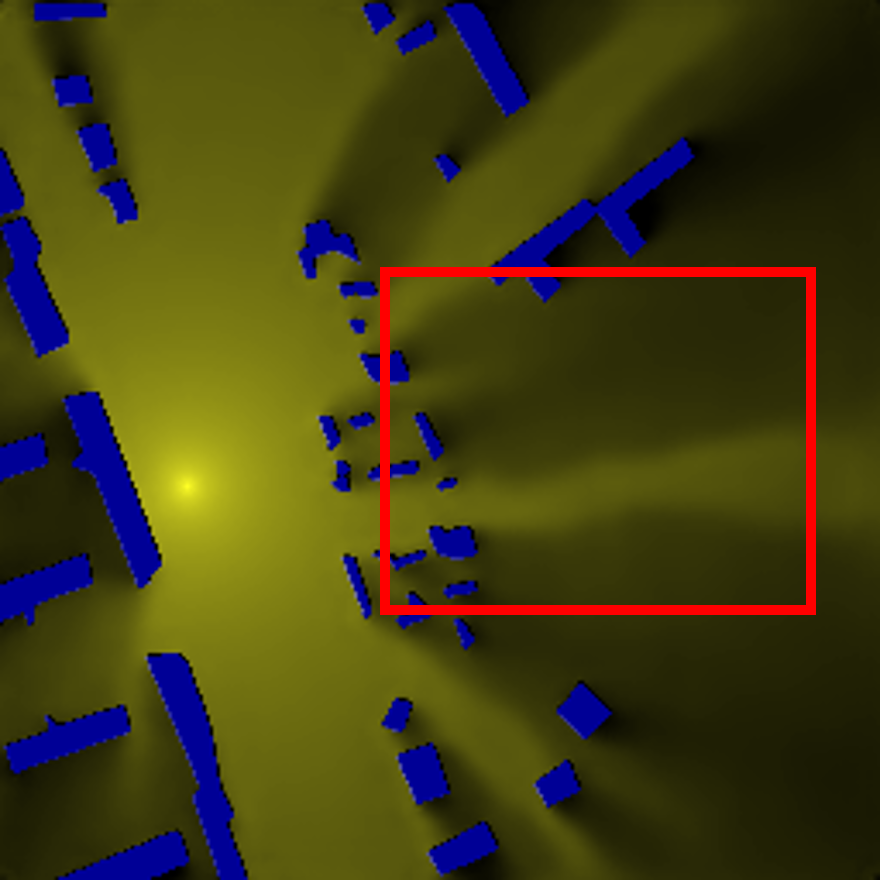}   \hspace{-4mm} &
                \includegraphics[width=0.18\linewidth]{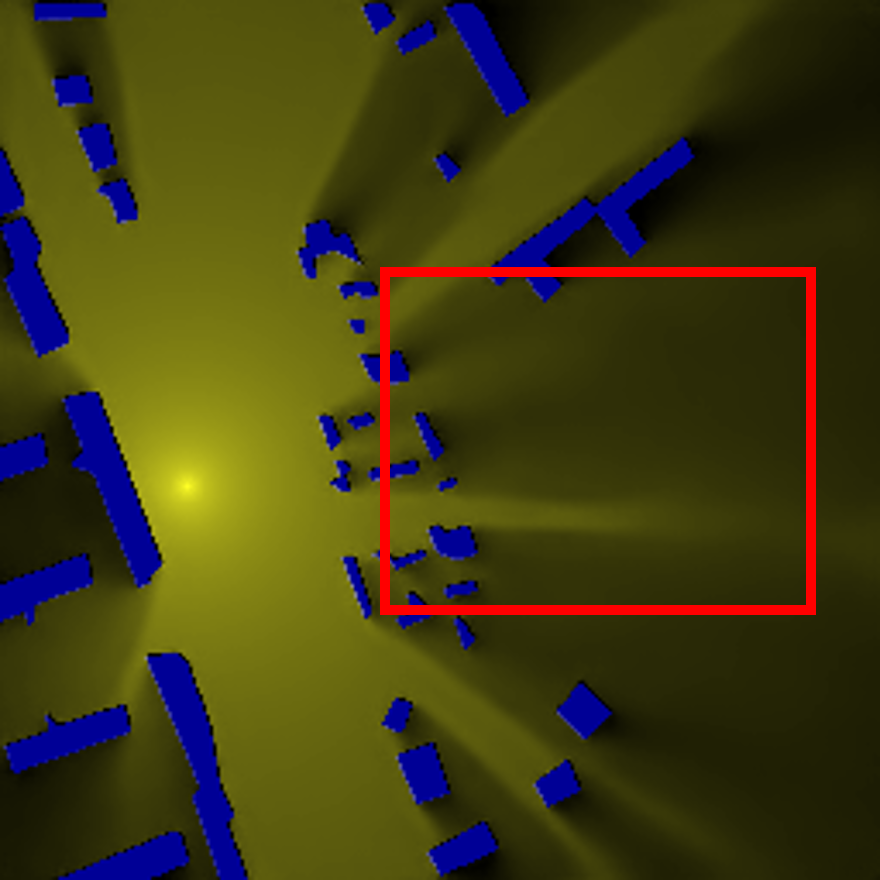}  \hspace{-4mm} &
                \includegraphics[width=0.18\linewidth]{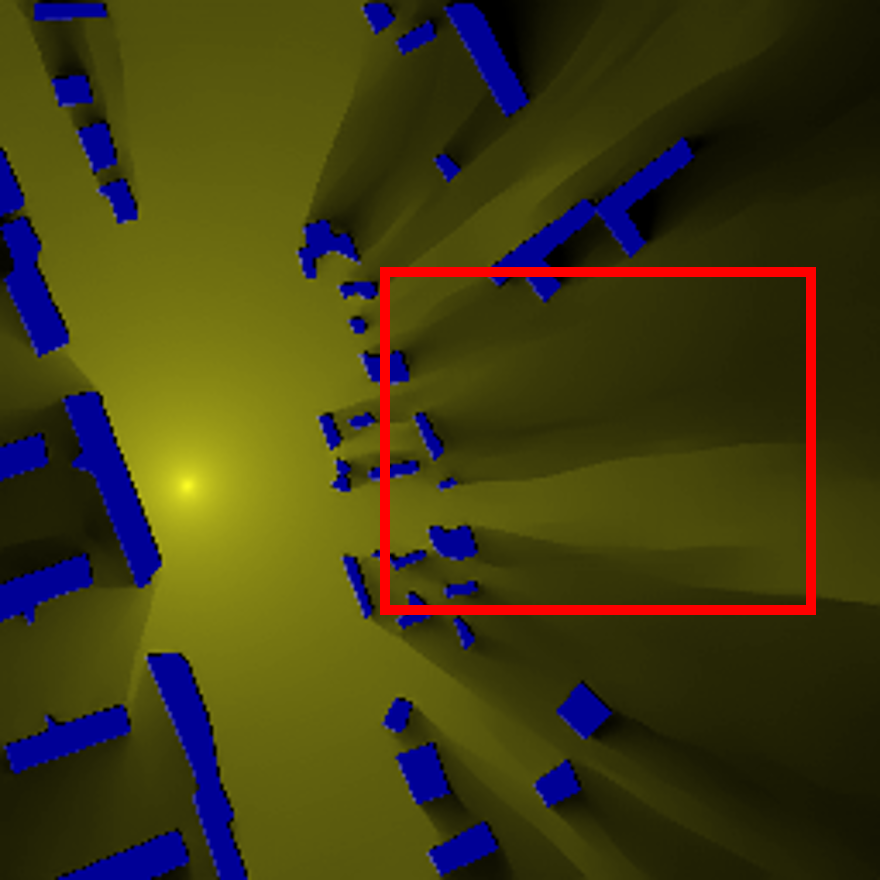}  \hspace{-4mm} &
                \includegraphics[width=0.18\linewidth]{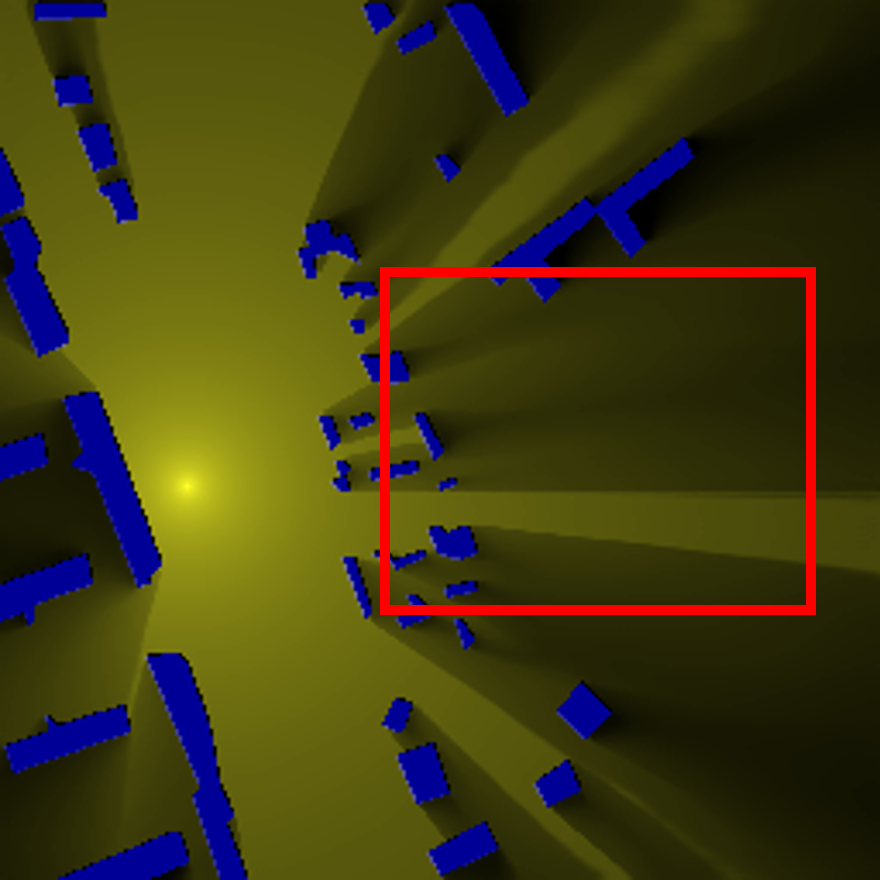} \hspace{-4mm} &
                \includegraphics[width=0.18\linewidth]{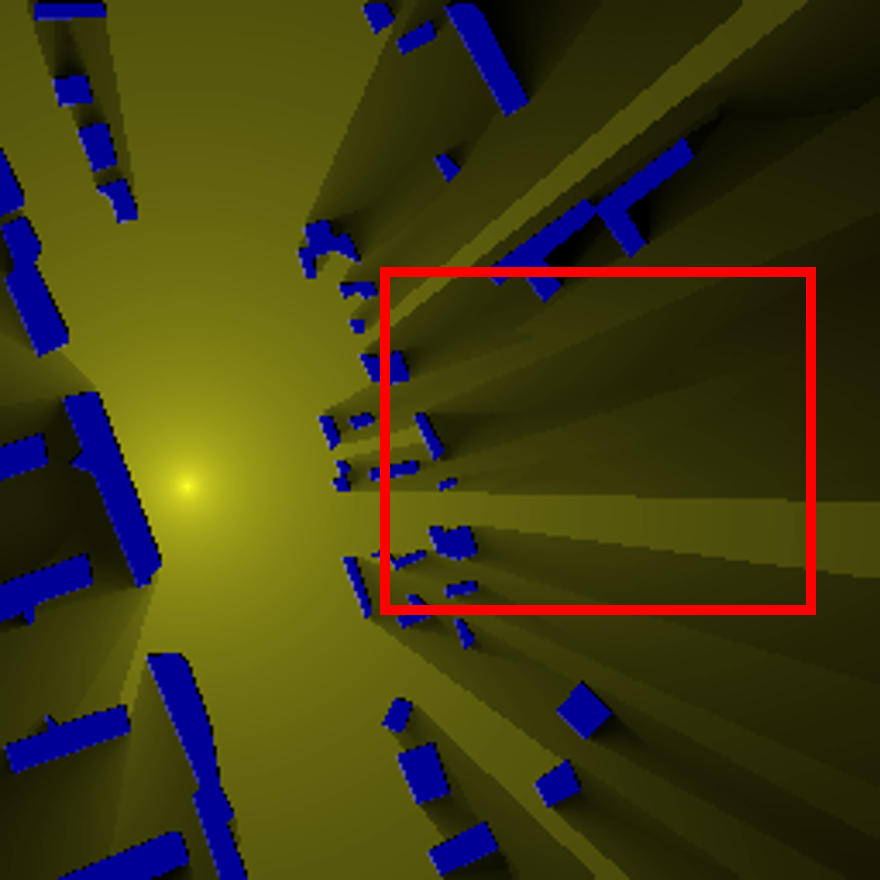}
            \end{tabular}
        \end{adjustbox} \\
        \vspace{1mm}

        \begin{adjustbox}{valign=t}
            \begin{tabular}{ccccc}
                \includegraphics[width=0.18\linewidth]{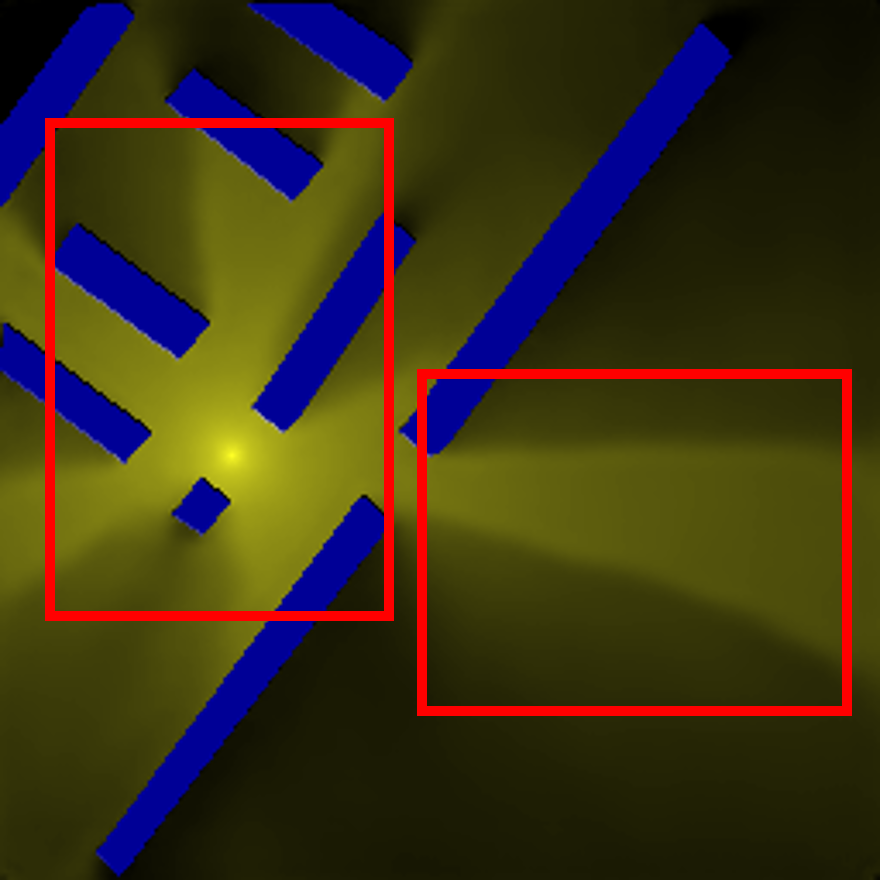}   \hspace{-4mm} &
                \includegraphics[width=0.18\linewidth]{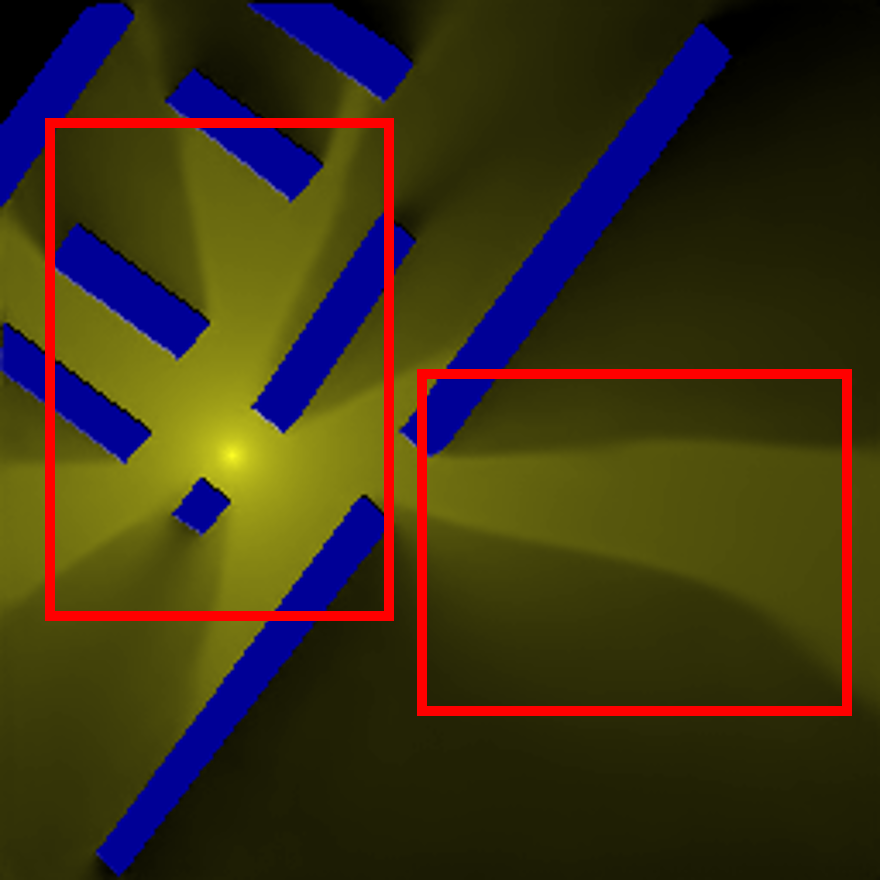}  \hspace{-4mm} &
                \includegraphics[width=0.18\linewidth]{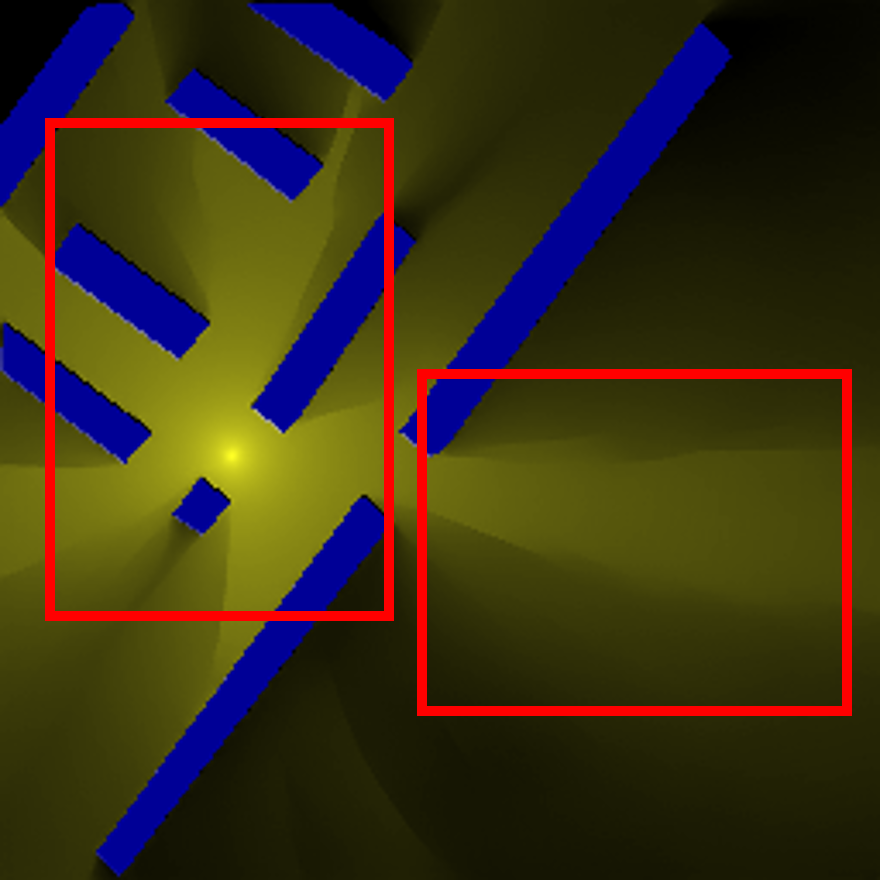}  \hspace{-4mm} &
                \includegraphics[width=0.18\linewidth]{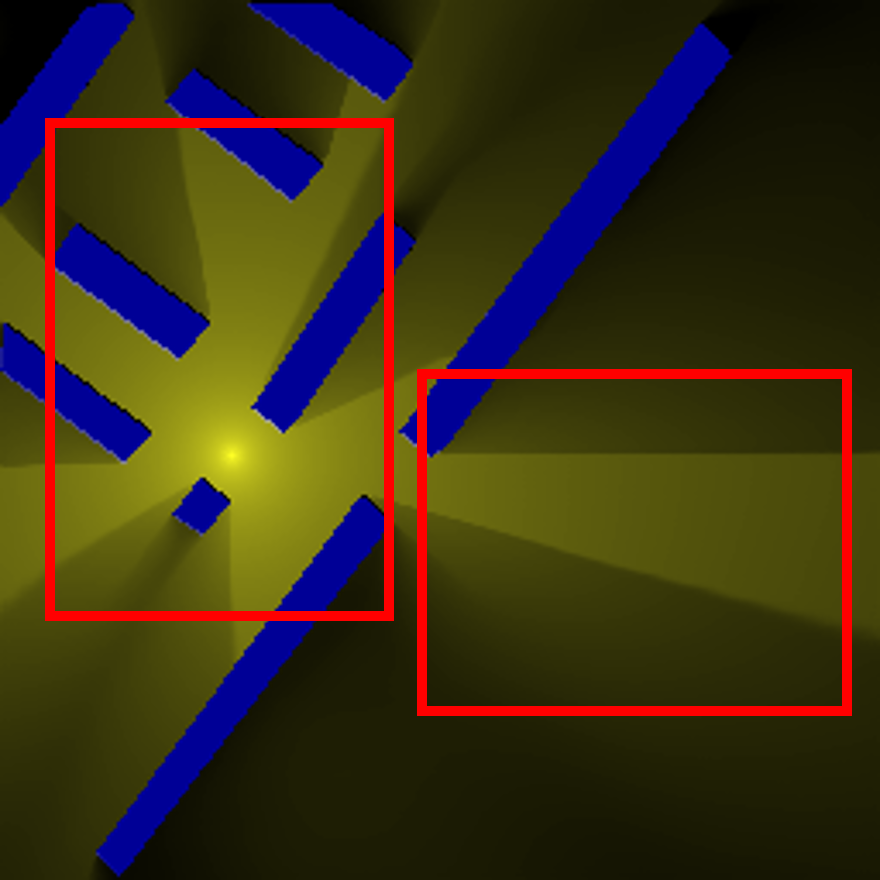} \hspace{-4mm} &
                \includegraphics[width=0.18\linewidth]{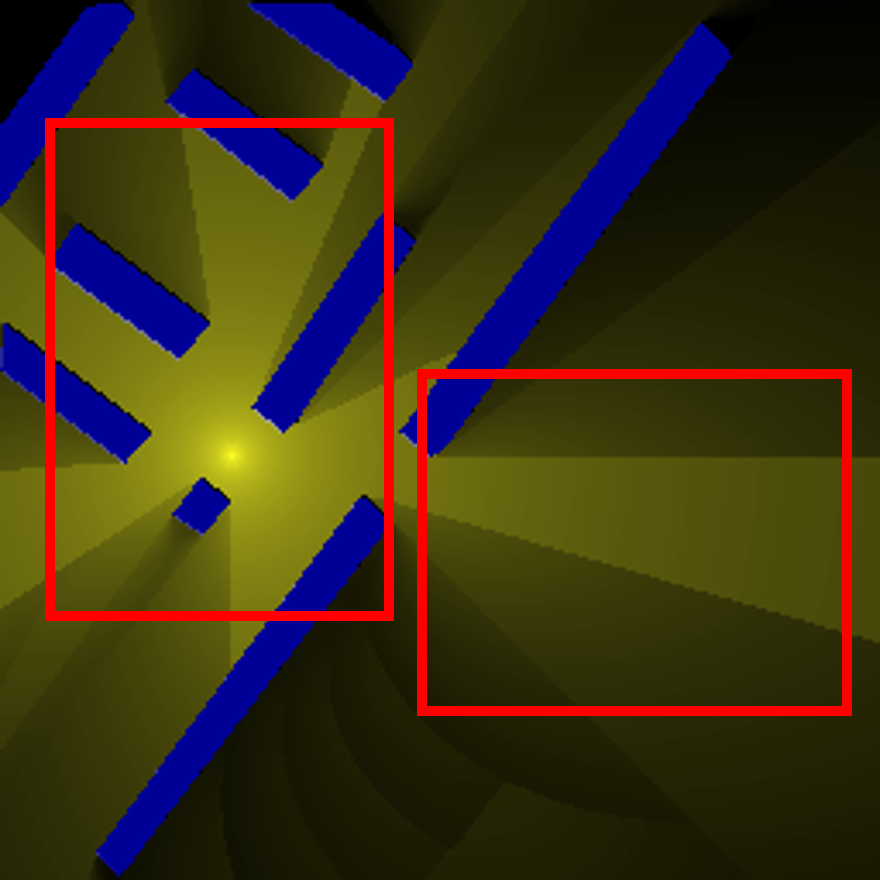}
            \end{tabular}
        \end{adjustbox} \\
        \vspace{1mm}

        \begin{adjustbox}{valign=t}
            \begin{tabular}{ccccc}
                \includegraphics[width=0.18\linewidth]{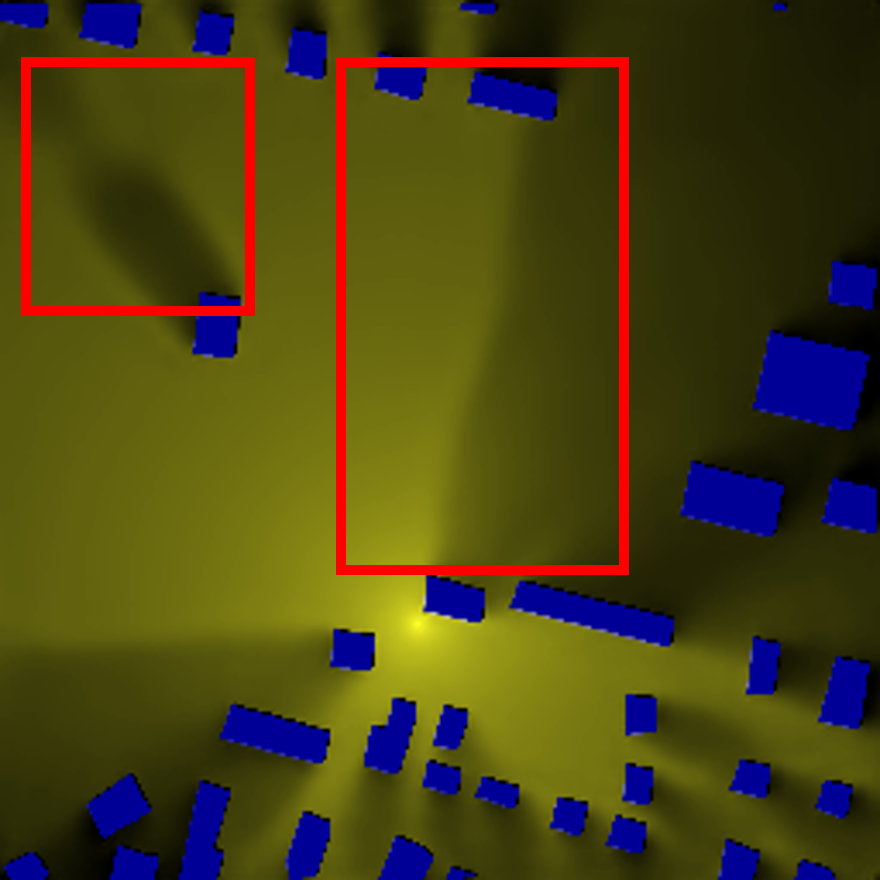}   \hspace{-4mm} &
                \includegraphics[width=0.18\linewidth]{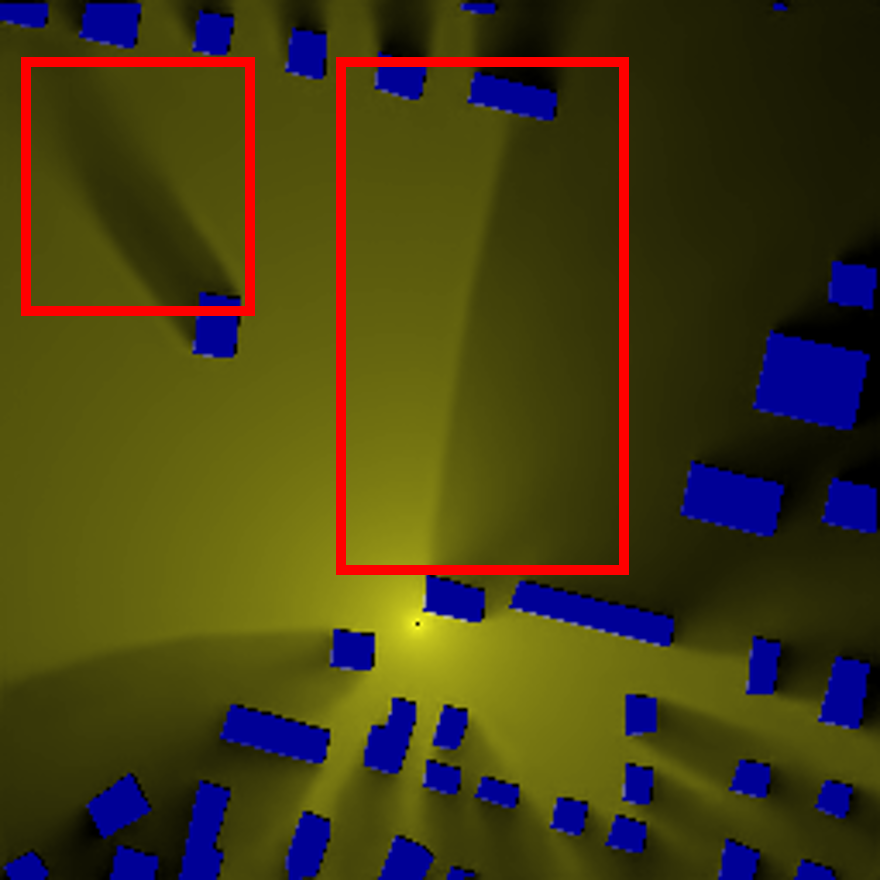}  \hspace{-4mm} &
                \includegraphics[width=0.18\linewidth]{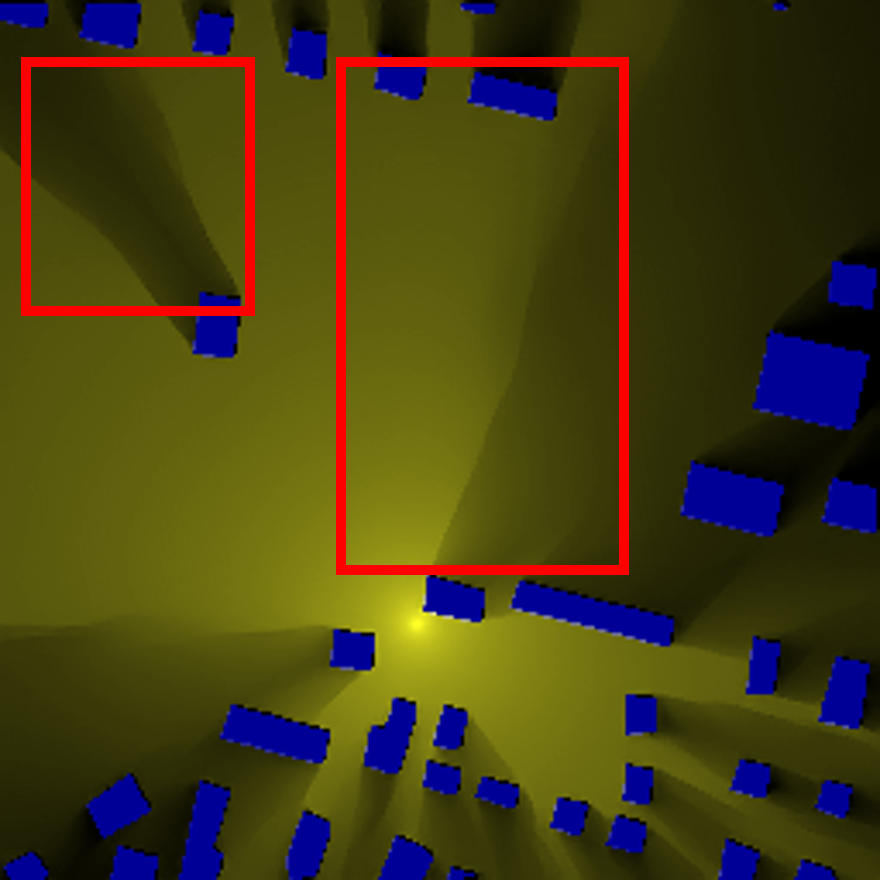}  \hspace{-4mm} &
                \includegraphics[width=0.18\linewidth]{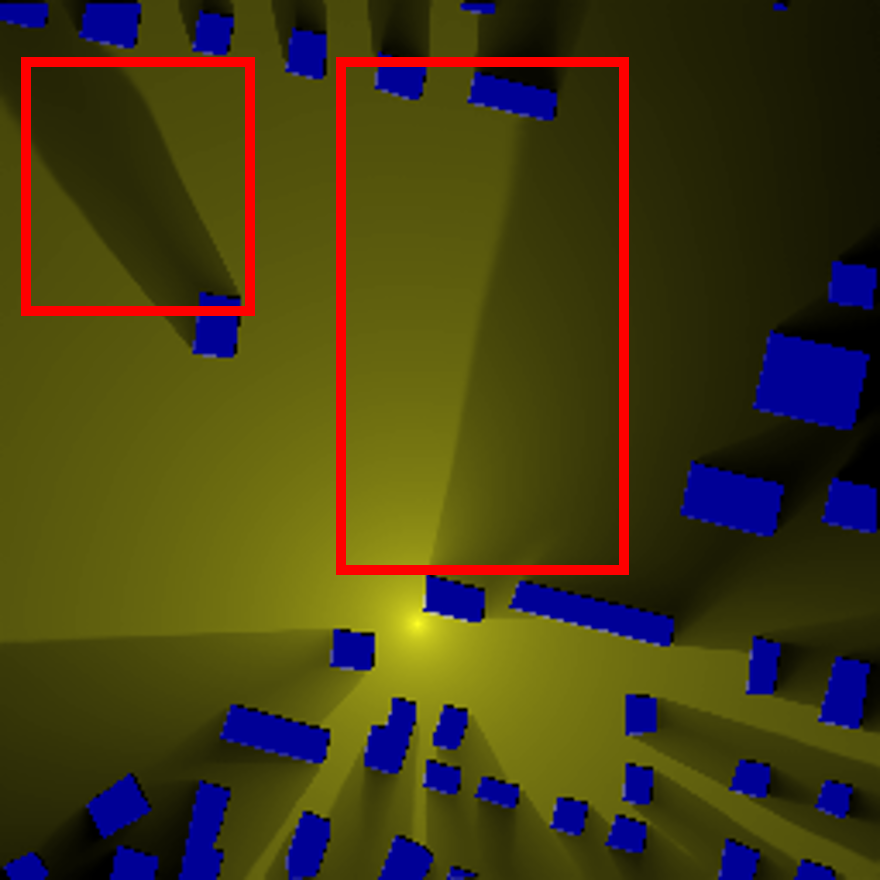} \hspace{-4mm} &
                \includegraphics[width=0.18\linewidth]{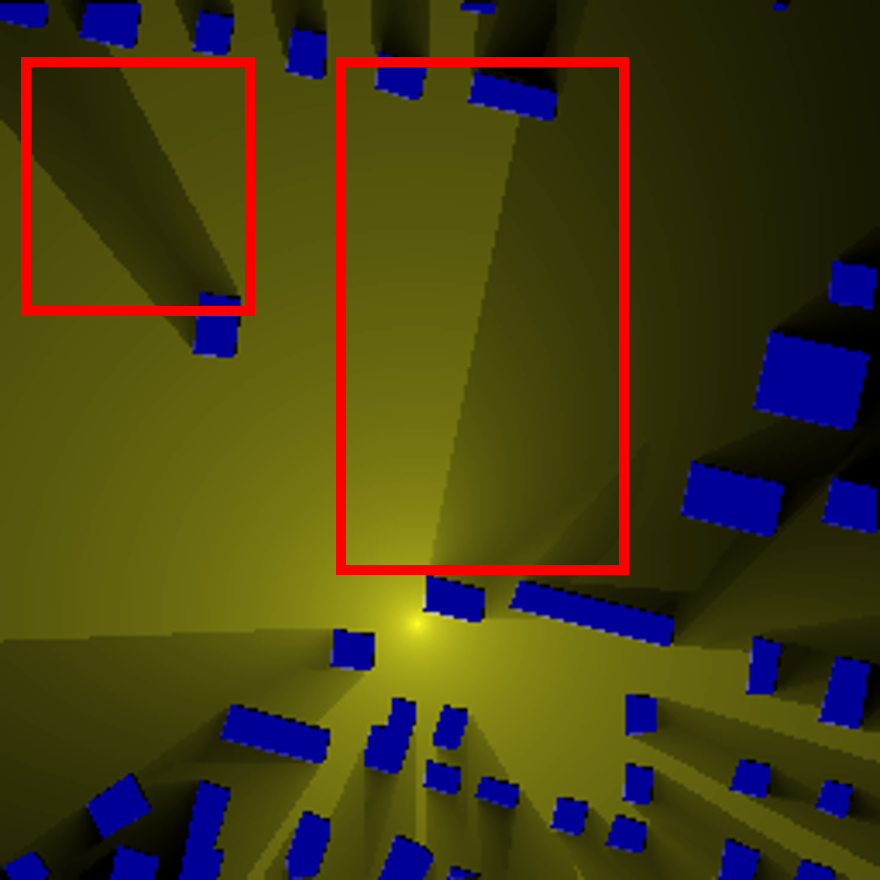} \\
                RME-GAN & RadioUNet & RadioDiff & RadioMamba (Ours) & Ground Truth
            \end{tabular}
        \end{adjustbox}
    \end{tabular}
    \caption{Qualitative comparison for SRM construction.}
    \label{fig:srm_qualitative_comparison}
    \vspace{-12pt}
\end{figure*}

\begin{figure*}[ht]
    \centering
    \captionsetup{font=small}
    \begin{tabular}{c}
        \begin{adjustbox}{valign=t}
            \begin{tabular}{ccccc}
                \includegraphics[width=0.18\linewidth]{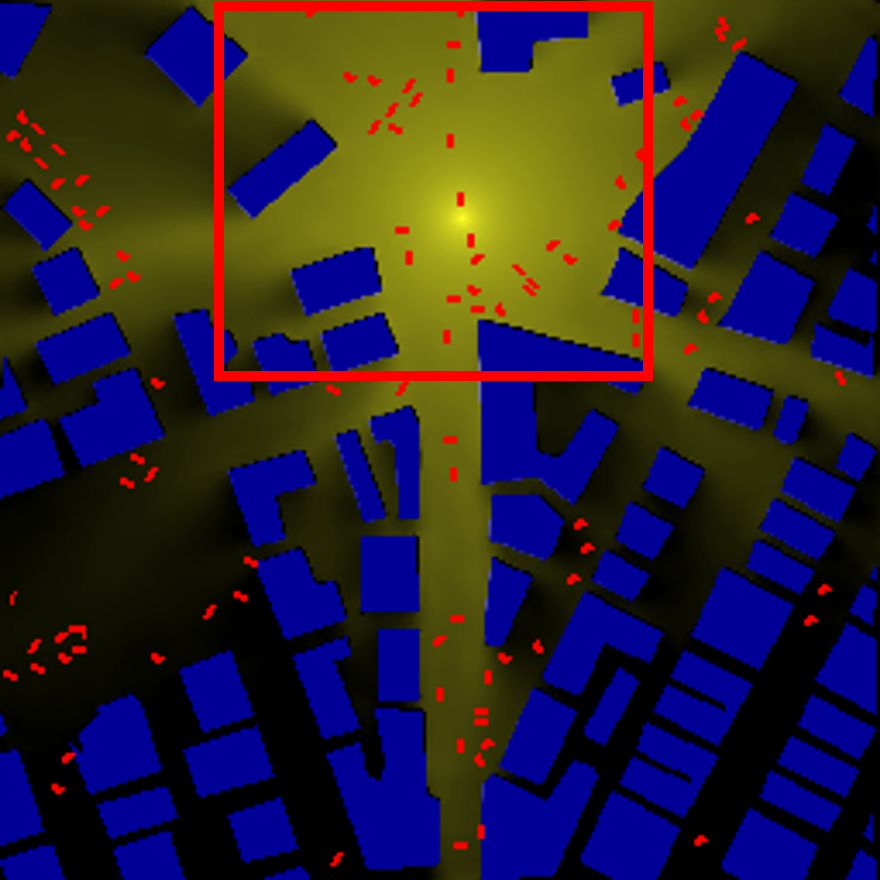}  \hspace{-4mm} &
                \includegraphics[width=0.18\linewidth]{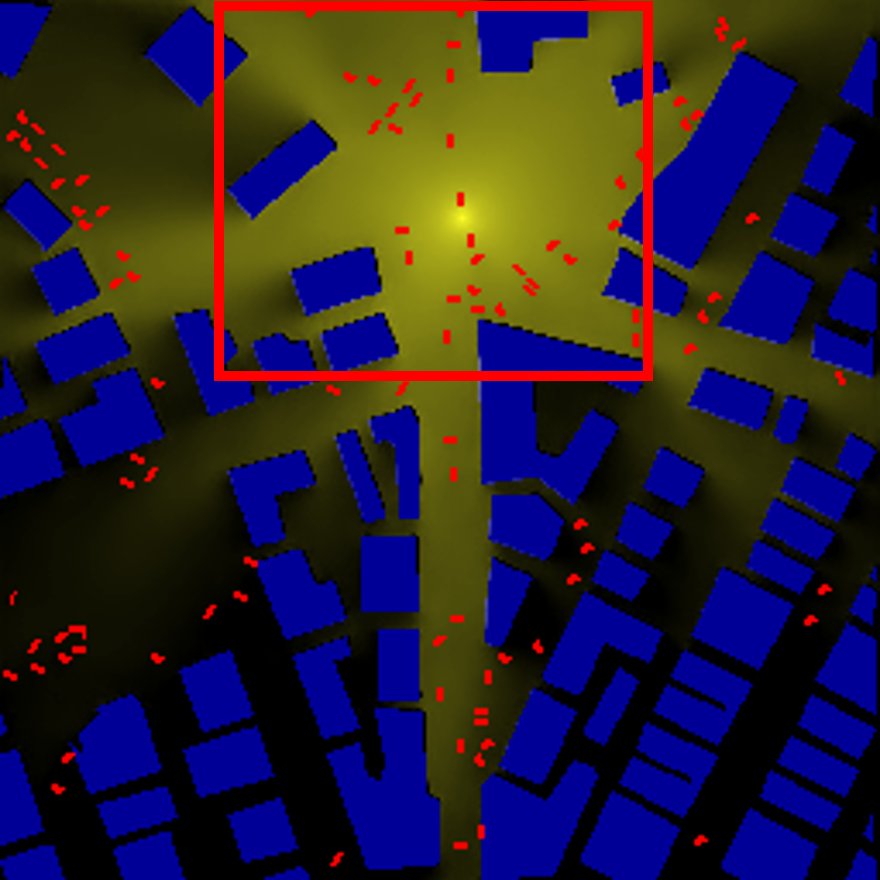}  \hspace{-4mm} &
                \includegraphics[width=0.18\linewidth]{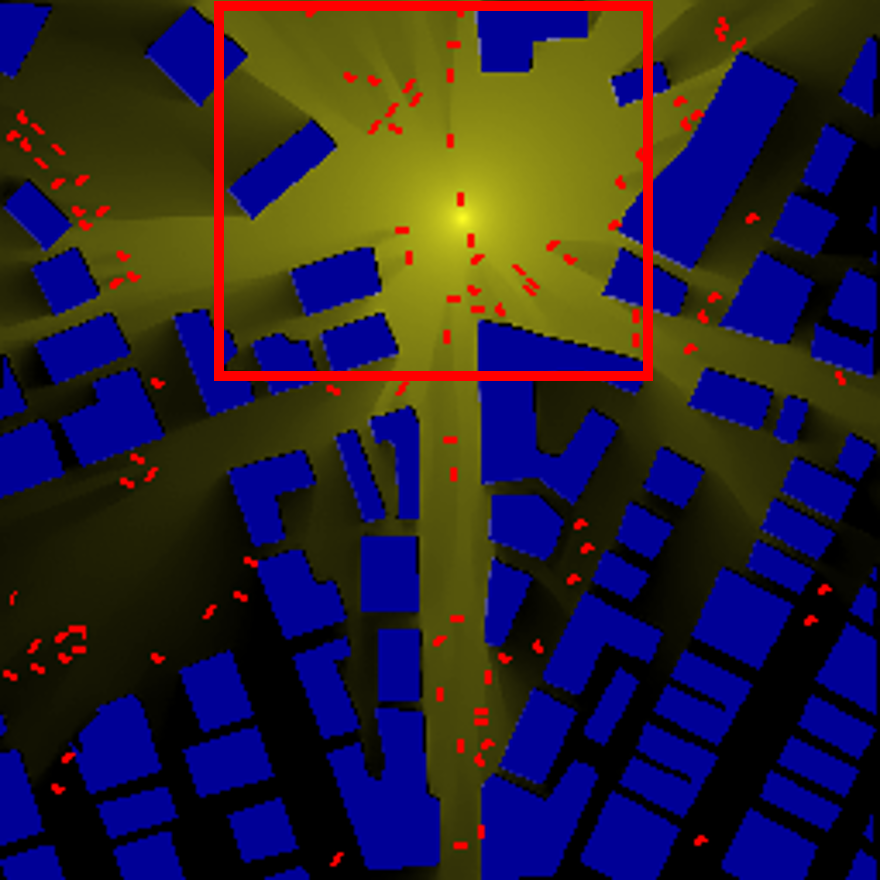}  \hspace{-4mm} &
                \includegraphics[width=0.18\linewidth]{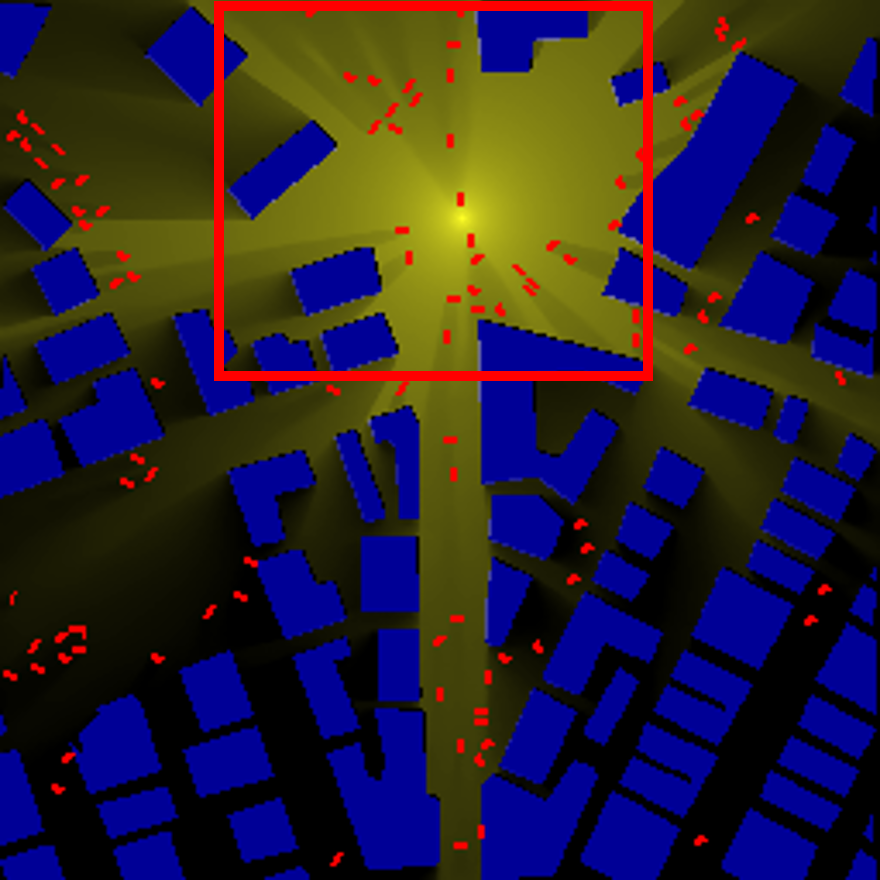} \hspace{-4mm} &
                \includegraphics[width=0.18\linewidth]{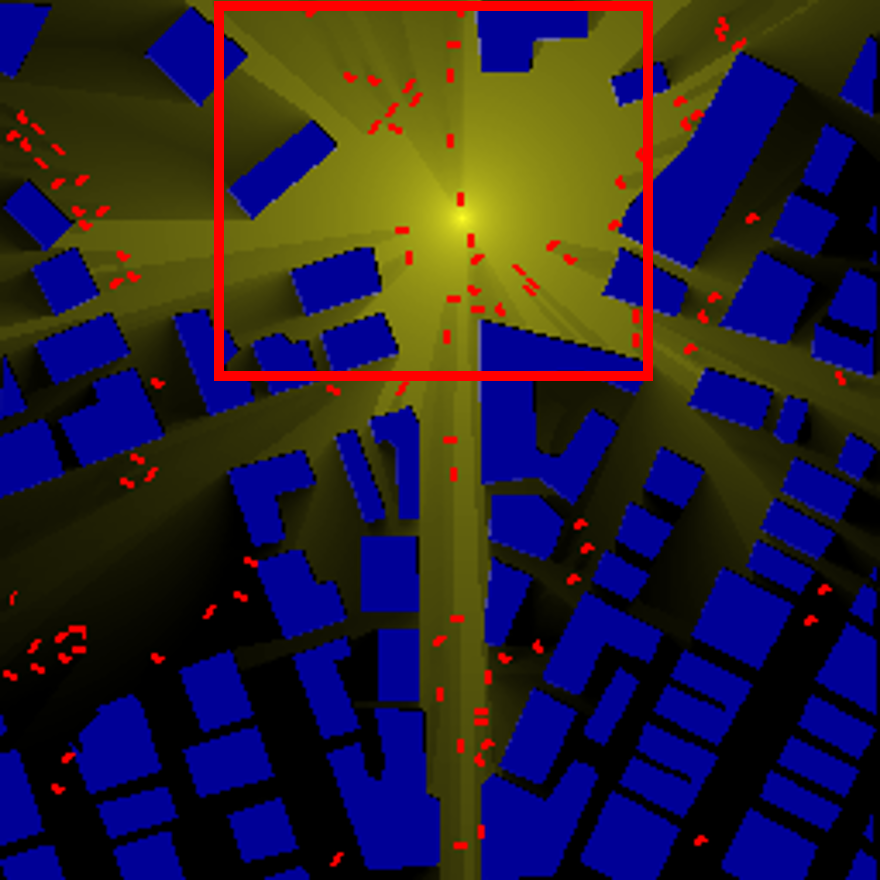}
            \end{tabular}
        \end{adjustbox} \\
        \vspace{1mm}

        \begin{adjustbox}{valign=t}
            \begin{tabular}{ccccc}
                \includegraphics[width=0.18\linewidth]{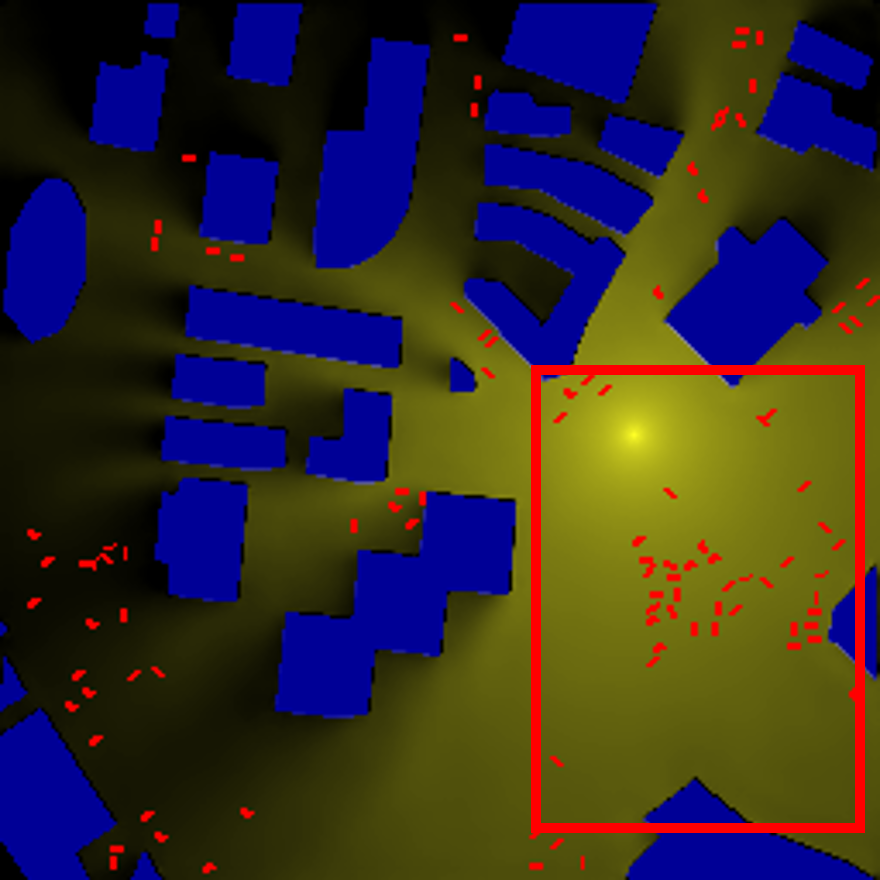}  \hspace{-4mm} &
                \includegraphics[width=0.18\linewidth]{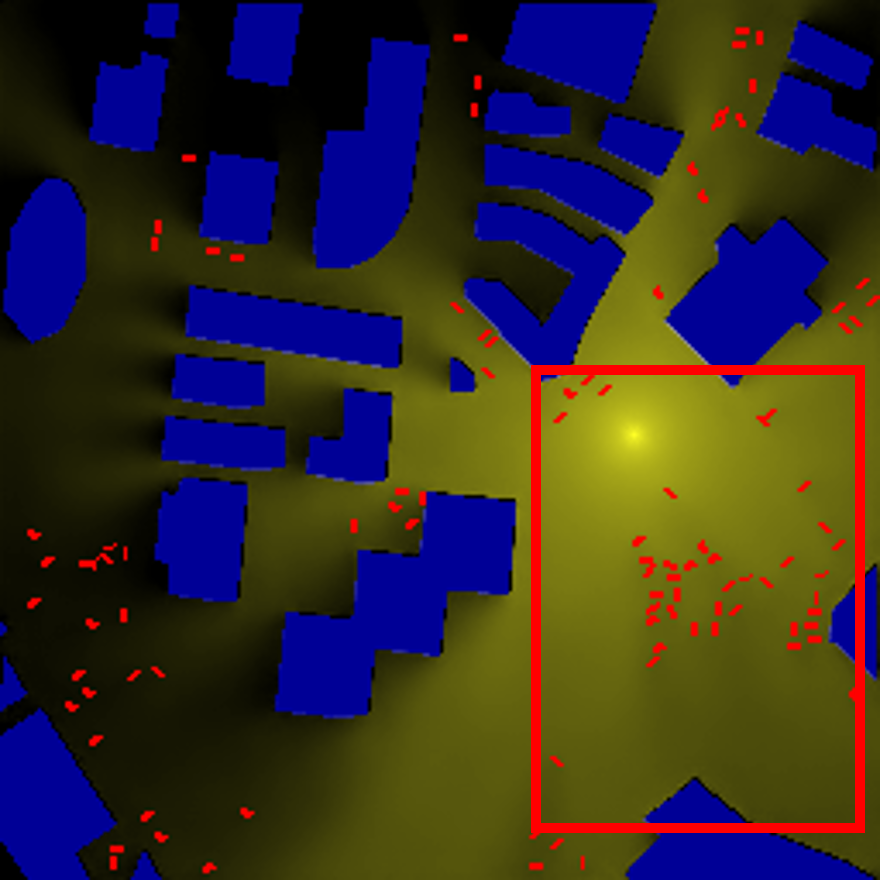}  \hspace{-4mm} &
                \includegraphics[width=0.18\linewidth]{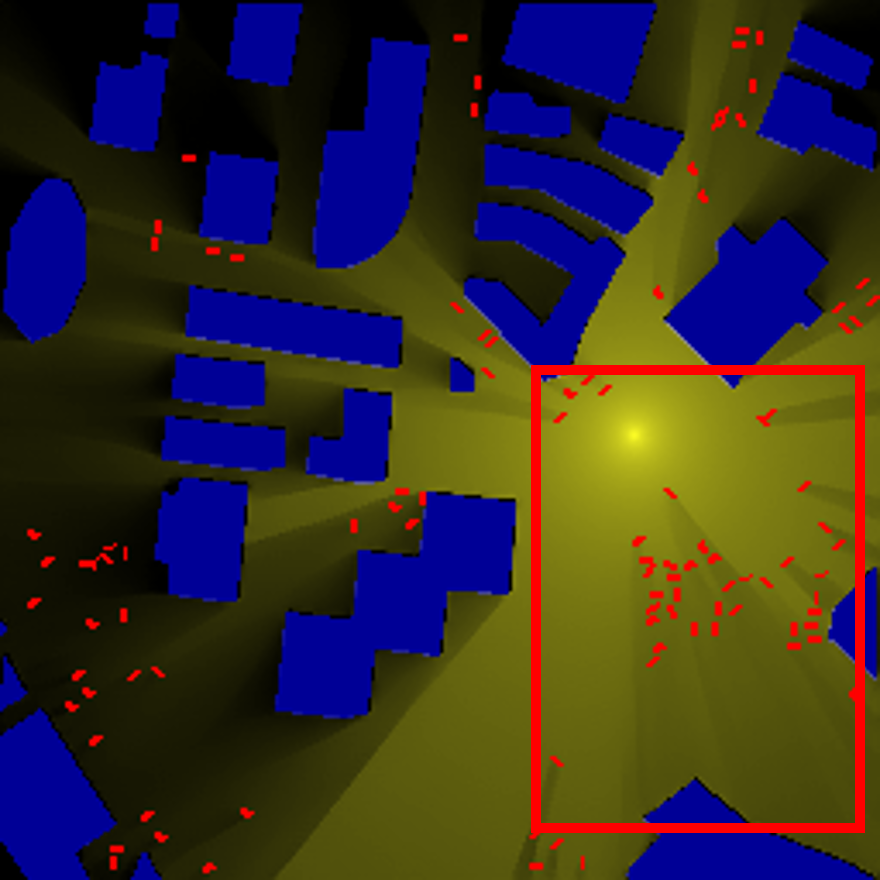}  \hspace{-4mm} &
                \includegraphics[width=0.18\linewidth]{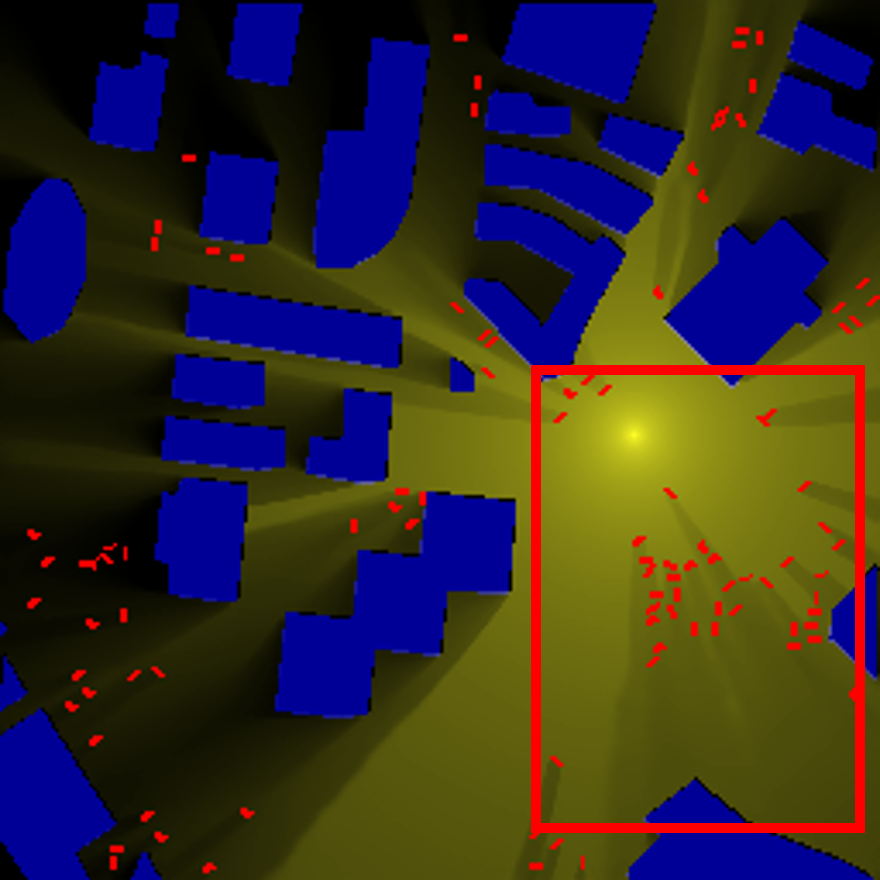} \hspace{-4mm} &
                \includegraphics[width=0.18\linewidth]{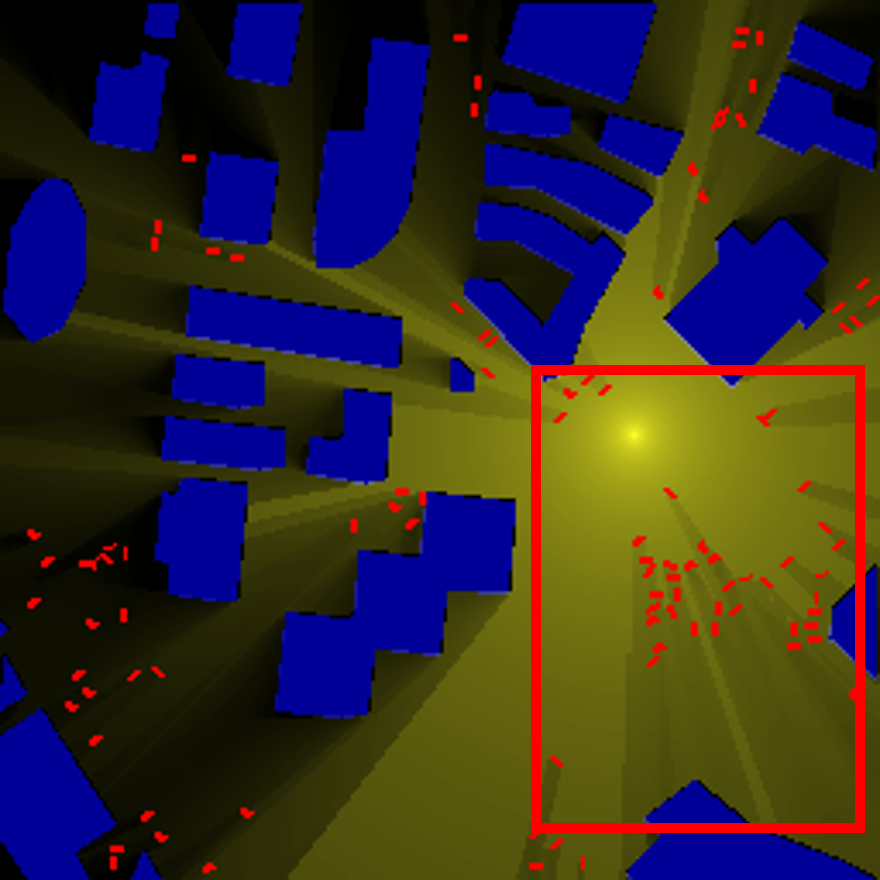}
            \end{tabular}
        \end{adjustbox} \\
        \vspace{1mm}

        \begin{adjustbox}{valign=t}
            \begin{tabular}{ccccc}
                \includegraphics[width=0.18\linewidth]{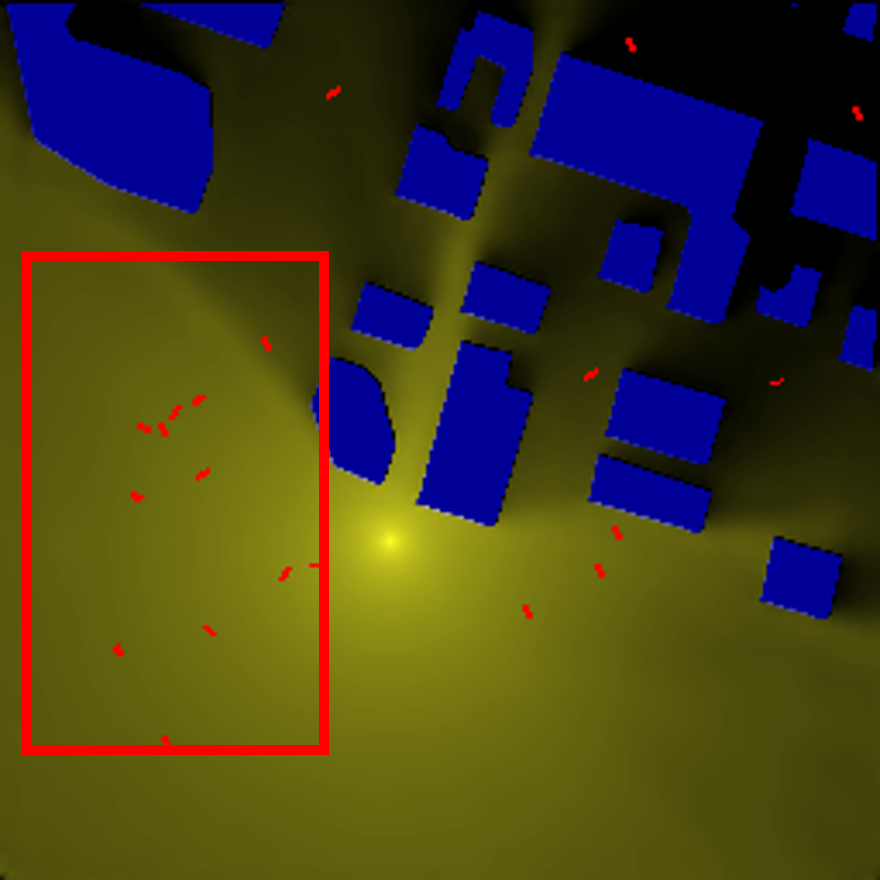}  \hspace{-4mm} &
                \includegraphics[width=0.18\linewidth]{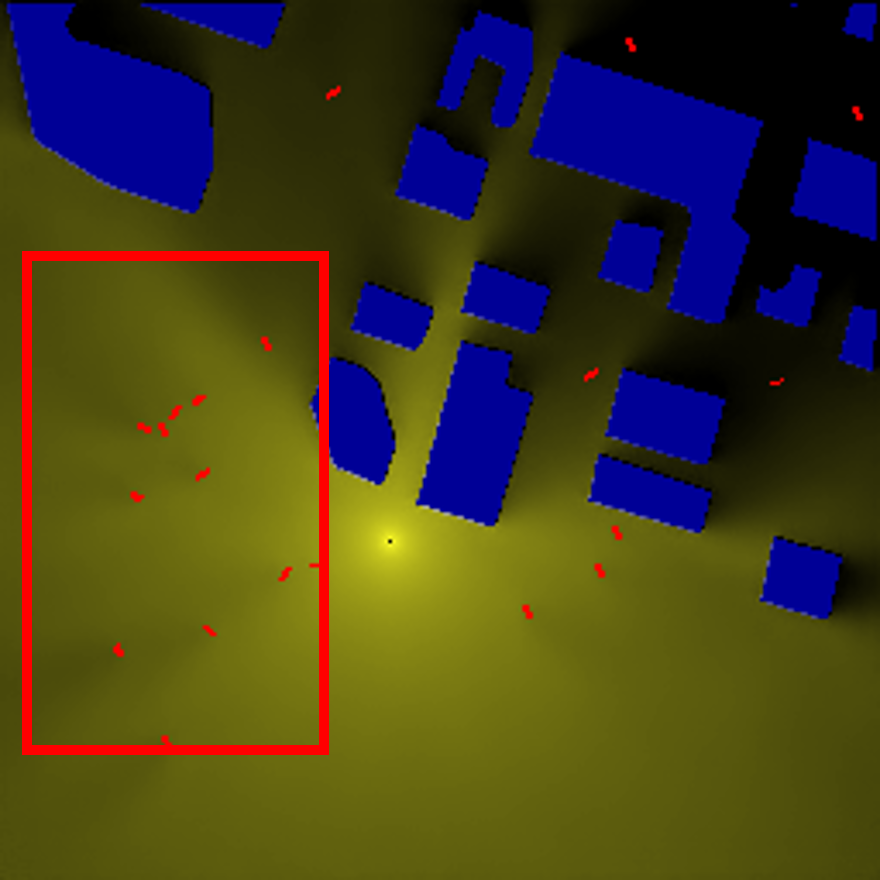}  \hspace{-4mm} &
                \includegraphics[width=0.18\linewidth]{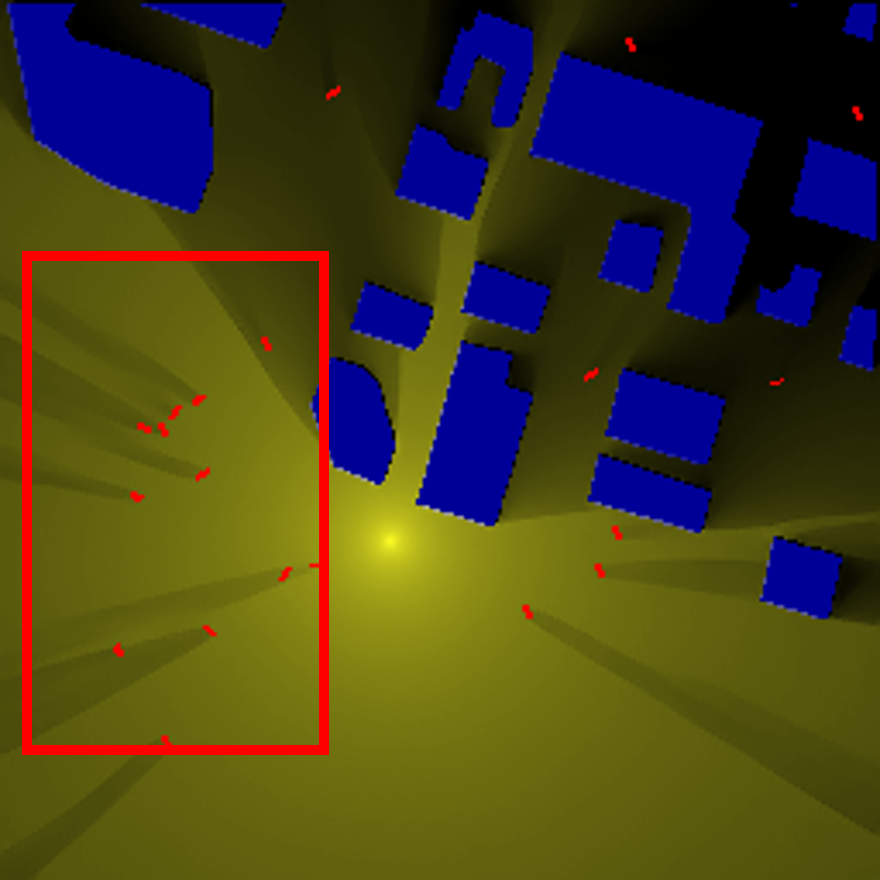}  \hspace{-4mm} &
                \includegraphics[width=0.18\linewidth]{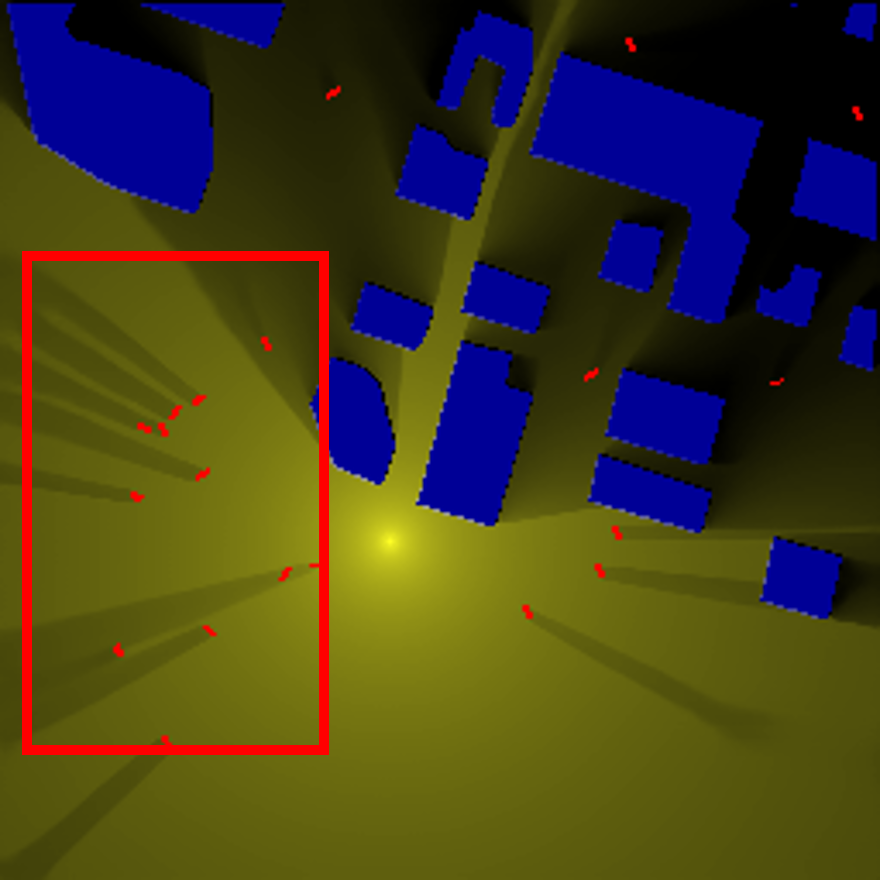} \hspace{-4mm} &
                \includegraphics[width=0.18\linewidth]{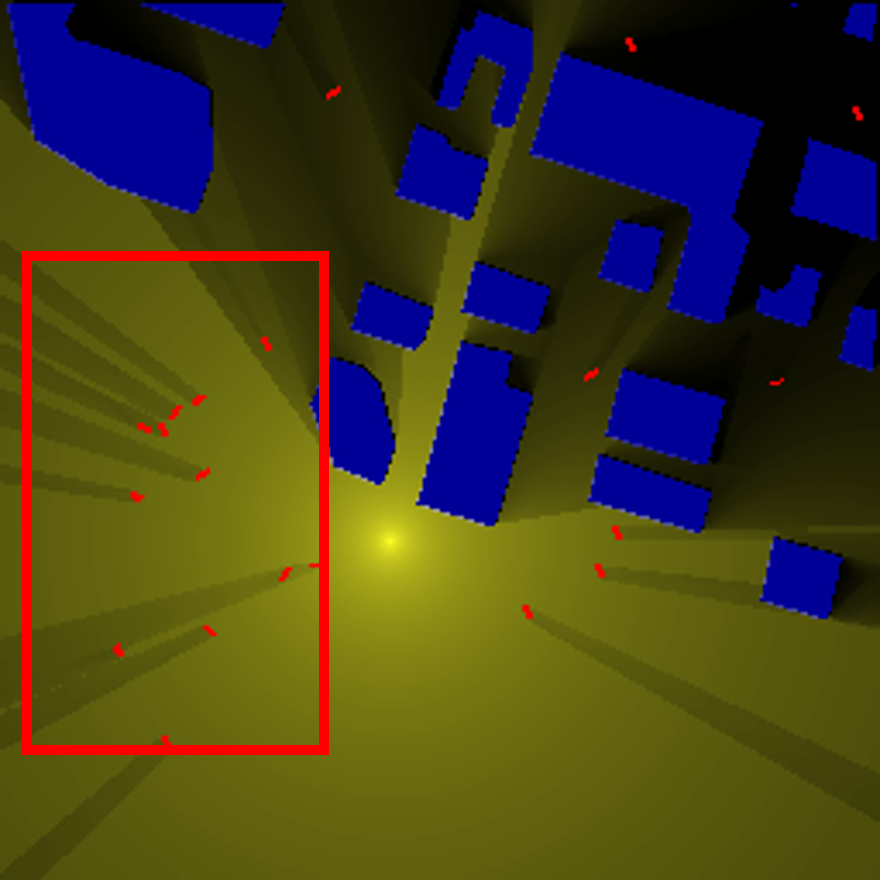}
            \end{tabular}
        \end{adjustbox} \\
        \vspace{1mm}

        \begin{adjustbox}{valign=t}
            \begin{tabular}{ccccc}
                \includegraphics[width=0.18\linewidth]{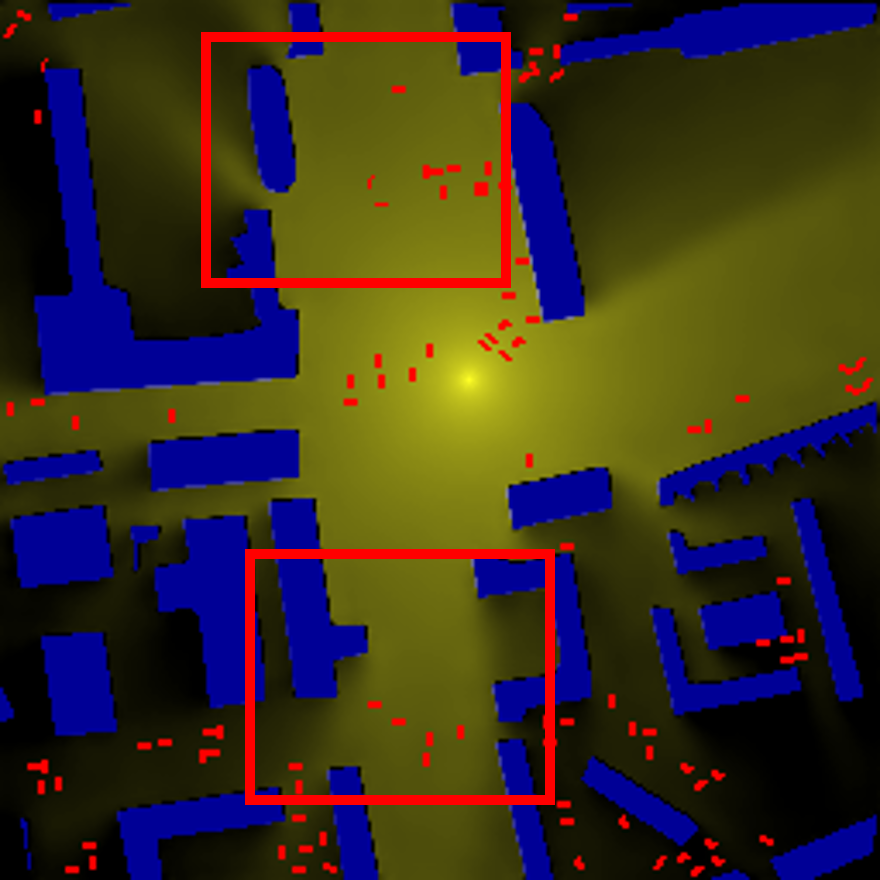}  \hspace{-4mm} &
                \includegraphics[width=0.18\linewidth]{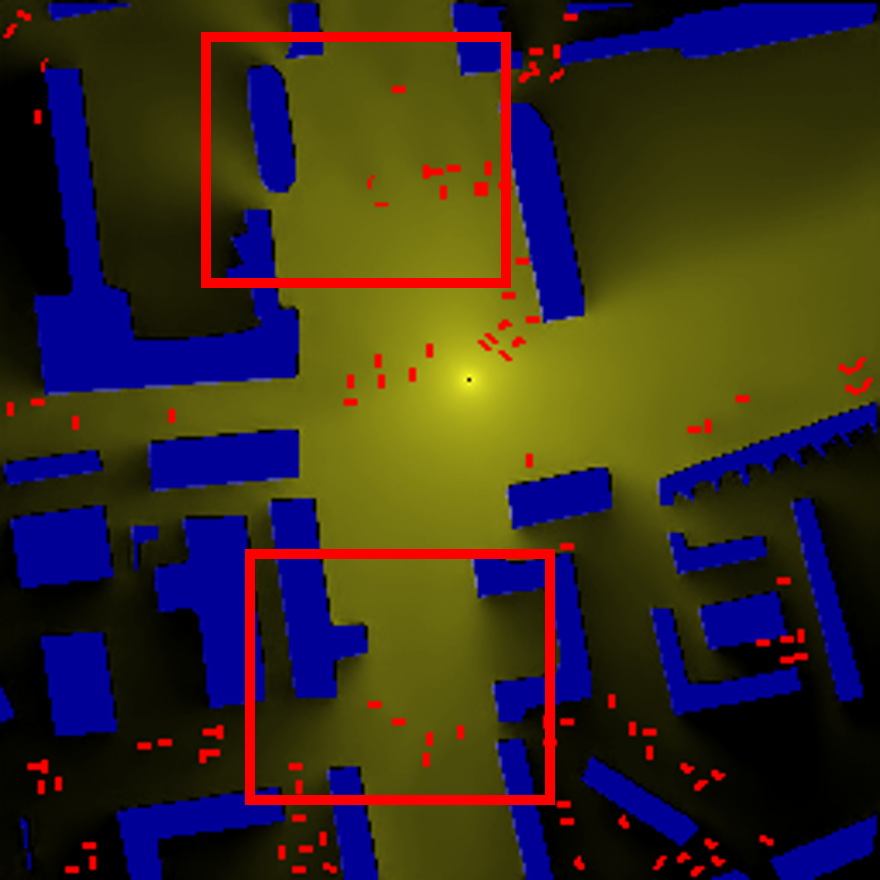}  \hspace{-4mm} &
                \includegraphics[width=0.18\linewidth]{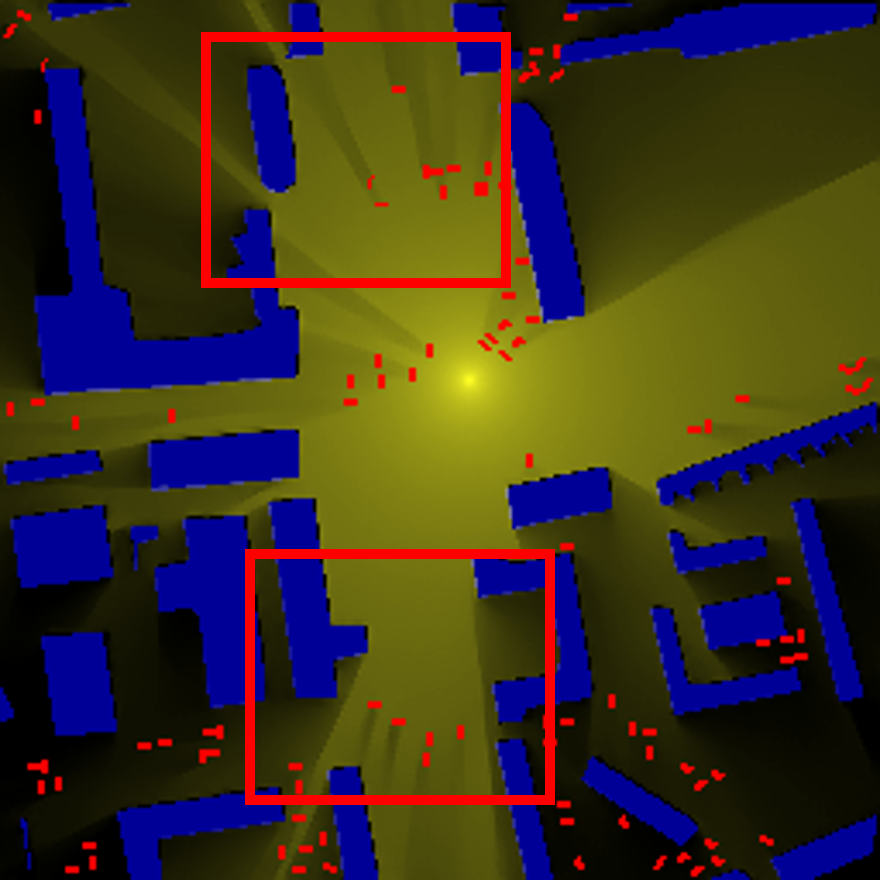}  \hspace{-4mm} &
                \includegraphics[width=0.18\linewidth]{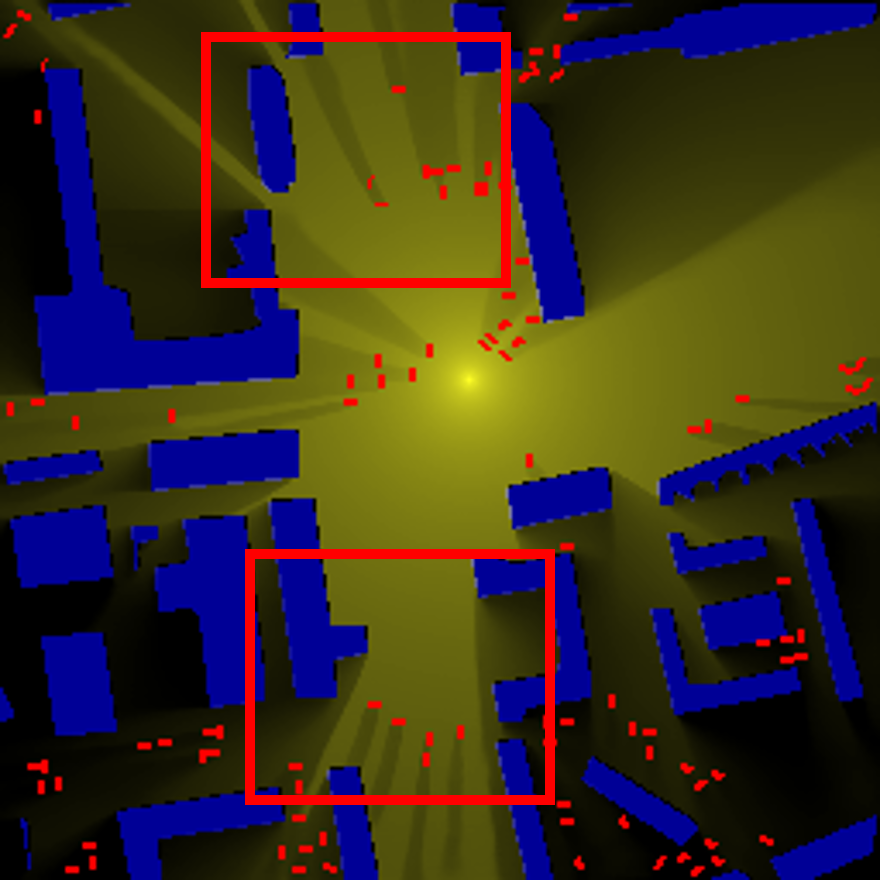} \hspace{-4mm} &
                \includegraphics[width=0.18\linewidth]{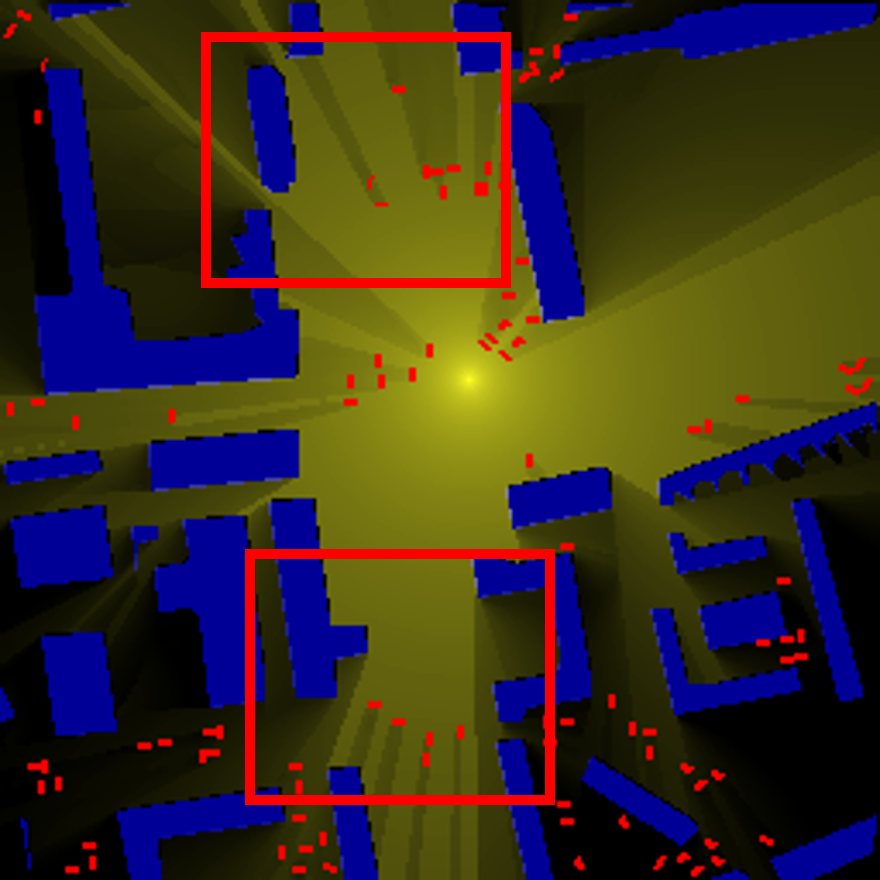} \\
                RME-GAN & RadioUNet & RadioDiff & RadioMamba (Ours) & Ground Truth
            \end{tabular}
        \end{adjustbox}
    \end{tabular}
    \caption{Qualitative comparison for DRM construction.}
    \label{fig:drm_qualitative_comparison}
    \vspace{-12pt}
\end{figure*}

\subsection{Experimental Setup}

\subsubsection{Dataset and Preprocessing}
We conduct all experiments on the publicly available \textbf{RadioMapSeer} dataset \cite{levie2021radiounet, wang2024radiodiff}, a standard benchmark for this task. The dataset consists of 700 maps from several cities, including Berlin and London, sourced from OpenStreetMap, with 80 different transmitter locations simulated for each map, totaling 56,000 unique scenarios. The maps are represented as $256 \times 256$ pixel images, where each pixel corresponds to a $1 \text{m} \times 1 \text{m}$ area. Building heights are fixed at 25 meters, and transmitter/receiver heights are set to 1.5 meters. The ground truth RMs are generated using a high-fidelity dominant path model (DPM) simulation. We evaluate performance on two sub-tasks:
\begin{itemize}
    \item \textbf{SRM construction}: Only static buildings are considered as obstacles. The input tensor comprises channels for the building layout and transmitter location.
    \item \textbf{DRM construction}: Both static buildings and transient obstacles, which are randomly generated vehicles along roads, are considered. An additional channel for dynamic obstacles is included in the input tensor.
\end{itemize}
We adhere to the standard data split of 550 maps for training, 50 for validation, and 100 for testing, ensuring no geographical overlap between sets. Input maps (buildings, transmitters, etc.) are converted to binary images. The output pathloss maps, originally in dB, are normalized to a grayscale range of [0, 1] for model training and evaluation.

\subsubsection{Baselines}
We compare our proposed RadioMamba with several SOTA baselines that represent different architectural philosophies to provide a comprehensive evaluation:
\begin{itemize}
    \item \textbf{RadioUNet} \cite{levie2021radiounet}: The influential CNN-based method. It is highly efficient but its accuracy is limited by the local receptive field of its convolutional operators.
    \item \textbf{RME-GAN} \cite{zhang2023rme}: A representative GAN-based model. It is efficient at inference but known for training instability and potential for artifacts.
    \item \textbf{RadioDiff} \cite{wang2024radiodiff}: The current SOTA in construction accuracy. This diffusion-based model generates high-fidelity maps via an iterative denoising process, but at the cost of being extremely parameter-heavy and having high inference latency.
\end{itemize}

\subsubsection{Implementation Details}
Our model was implemented using PyTorch and trained on NVIDIA A40 GPUs. We utilized the AdamW optimizer with an initial learning rate of $9 \times 10^{-4}$ and a weight decay of $10^{-4}$. A cosine annealing scheduler was employed to manage the learning rate. To accelerate training and reduce memory footprint, we used 16-bit mixed-precision. The final RadioMamba model has approximately 8.6 million parameters, highlighting its lightweight nature.

\subsubsection{Loss Function}
To guide the network towards producing physically plausible and perceptually high-quality maps, we employ a composite loss function. The total loss $\mathcal{L}_{\text{total}}$ is a weighted sum of four components as follows.
\begin{equation}
\mathcal{L}_{\text{total}} = w_1\mathcal{L}_{\text{L1}} + w_2\mathcal{L}_{\text{MSE}} + w_3\mathcal{L}_{\text{SSIM}} + w_4\mathcal{L}_{\text{Grad}}.
\end{equation}
The components are as follows, where $\hat{\mathbf{P}}$ is the predicted radio map and $\mathbf{P}$ is the ground truth.
\begin{itemize}
    \item \textbf{Mean absolute error ($\mathcal{L}_{\text{L1}}$)} and \textbf{mean squared error ($\mathcal{L}_{\text{MSE}}$)}: These standard pixel-wise losses ensure numerical accuracy.
    \item \textbf{Structural similarity loss ($\mathcal{L}_{\text{SSIM}}$)}: Formulated as $1 - \text{SSIM}(\hat{\mathbf{P}}, \mathbf{P})$, this loss enhances perceptual quality by focusing on structural information.
    \item \textbf{Gradient loss ($\mathcal{L}_{\text{Grad}}$)}: To enforce sharp edges and preserve fine details, we compute the L1 loss between the gradients of the prediction and the target, obtained via a Sobel filter: $\mathcal{L}_{\text{Grad}} = ||\nabla \hat{\mathbf{P}} - \nabla \mathbf{P}||_1$. This is crucial for accurately rendering diffraction patterns.
\end{itemize}
Through empirical validation, the weights were set to $w_1=0.4, w_2=0.1, w_3=0.2, w_4=0.3$ to achieve a balanced optimization.

\subsubsection{Evaluation Metrics}
We provide a comprehensive evaluation using four metrics for construction quality and three for computational efficiency.
\begin{itemize}
    \item \textbf{Accuracy Metrics}: normalized mean square error (NMSE) and root mean square error (RMSE).
    \item \textbf{Perceptual Metrics}: structural similarity index measure (SSIM) and peak signal-to-noise ratio (PSNR, in dB).
    \item \textbf{Efficiency Metrics}: inference time (seconds per map), model parameters (millions), and GPU memory usage (MB). Lower is better for all three.
\end{itemize}

\subsection{Results and Analysis}
In this section, we analyze RadioMamba's performance, comparing it to other methods in terms of accuracy and efficiency. Our results show high accuracy and efficiency, addressing the trade-off between the two.

\subsubsection{SRM construction Results}
The SRM task serves as a fundamental benchmark for a model's ability to learn propagation physics. In Table \ref{tab:combined_rm_results}, the best-performing metrics are highlighted in bold red, while the second-best are underlined in blue.

As shown in Table \ref{tab:combined_rm_results}, RadioMamba achieves lower error and higher similarity scores compared to all baselines, including the RadioDiff model. The reduction in NMSE from 0.0072 (RadioDiff) to 0.0050 represents a \textbf{30.5\% improvement}, while the RMSE is reduced by \textbf{17.1\%}. This performance gain can be attributed to the model's hybrid architectural design. The pathloss at any point is governed by global phenomena, such as large-scale shadowing from distant building clusters. The Mamba branch, with its global receptive field, is equipped to model these long-range interactions. Simultaneously, the convolutional branch captures fine-grained, local diffraction patterns. By combining these two capabilities, RadioMamba develops a more holistic and physically consistent understanding of the radio environment than models reliant on purely local (CNNs) or purely iterative (Diffusion) mechanisms.

\subsubsection{DRM construction Results}
The DRM task introduces additional complexity by adding transient obstacles, simulating a more realistic urban environment. The results are presented in Table \ref{tab:combined_rm_results}.

As expected, the performance of all models degrades in this more challenging scenario. However, RadioMamba maintains its lead, once again outperforming all competitors by a wide margin. It achieves an NMSE of 0.0063, a \textbf{30\% improvement} over RadioDiff's 0.0090. This result demonstrates the model's adaptability. The model effectively leverages its global context awareness (via Mamba) to account for the large-scale changes in signal clutter introduced by dynamic obstacles, while its local feature extraction (via CNN) precisely delineates the new, sharp shadows they cast. This confirms that the benefits of our hybrid design are not limited to static cases but are equally potent in complex, changing environments.

\subsubsection{Efficiency and Complexity}
While accuracy is crucial, practical deployment in 6G systems is ultimately dictated by efficiency. Table \ref{tab:efficiency} provides a comparison of model complexity and resource consumption, where the percentages indicate the performance improvement of our model relative to RadioDiff.

RadioMamba performs inference in just \textbf{28 milliseconds}, a speed that is \textbf{nearly 20 times faster} than RadioDiff's 553 milliseconds. This latency is well within the stringent requirements for real-time applications like network control loops or UAV navigation. In contrast, RadioDiff's half-second delay relegates it to offline analysis tasks.

Furthermore, RadioMamba achieves its SOTA accuracy with only \textbf{8.6 million parameters}, a \textbf{97.1\% reduction} compared to RadioDiff's 297.74 million. This reduction in model size, along with its lower memory footprint (808 MB vs. 2067 MB), makes RadioMamba a prime candidate for deployment on edge devices with limited computational resources. It challenges the notion that top-tier accuracy must come at the cost of massive model size and high latency.

\subsubsection{Qualitative Comparison}
Quantitative metrics are crucial, but a visual inspection of the generated maps provides deeper insight into model performance. Figures \ref{fig:srm_qualitative_comparison} and \ref{fig:drm_qualitative_comparison} display sample outputs from all models against the ground truth.

The qualitative results are consistent with our quantitative findings. The maps generated by RadioMamba are visually more accurate, exhibiting a close resemblance to the ground truth. Observe in Figure \ref{fig:srm_qualitative_comparison} (e.g., row 4), how RadioMamba accurately captures the subtle gradations of signal decay in open spaces and the sharp, well-defined reflection patterns behind buildings. In contrast, RME-GAN and RadioUNet's output appears smoothed and lacks fine detail, a direct consequence of its limited receptive field. Even RadioDiff, while generally accurate, tends to slightly blur the sharpest transitions. In the challenging DRM scenarios (Figure \ref{fig:drm_qualitative_comparison}), RadioMamba's ability to render clean, crisp shadow edges cast by both static and dynamic obstacles is particularly evident. This visual fidelity confirms its advanced understanding of the underlying wave propagation physics.

\subsection{Ablation Study}
To validate our design and analyze the contributions of each component, we conduct a comprehensive series of ablation studies. Our investigation is structured into three parts: analyzing the core MambaConvBlock, justifying our choice of efficient convolutions, and evaluating the impact of the loss function components.

\subsubsection{Component Analysis of the MambaConvBlock}
The central innovation of RadioMamba is the synergistic MambaConvBlock, which combines a global Mamba branch and a local convolution branch. To demonstrate that this hybrid approach is better to using either component in isolation, we compare our full model against two variants: first, a \textbf{Conv-Only} baseline, where the Mamba branch is removed, and second, a \textbf{Mamba-Only} baseline, where the convolution branch is removed.

\begin{figure}[ht]
    \centering
    \captionsetup{font=small}
    \begin{tabular}{cc}
        \includegraphics[width=0.35\columnwidth]{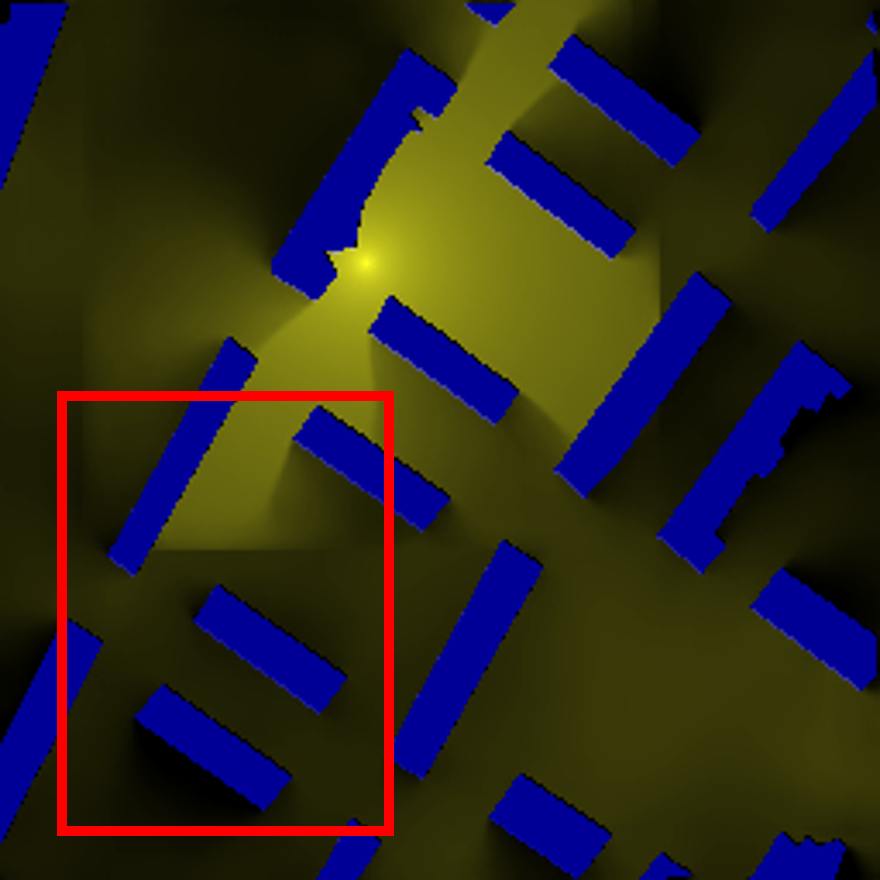} & 
        \includegraphics[width=0.35\columnwidth]{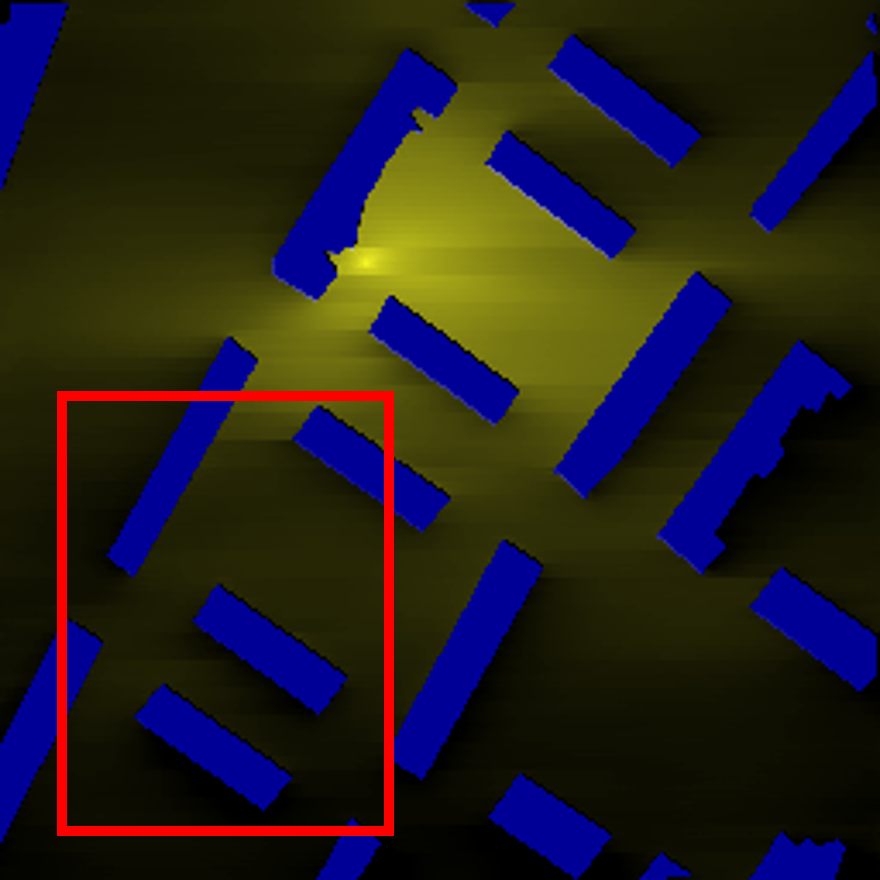} \\
        \small (a) Conv-Only & \small (b) Mamba-Only \\[6pt]
        \includegraphics[width=0.35\columnwidth]{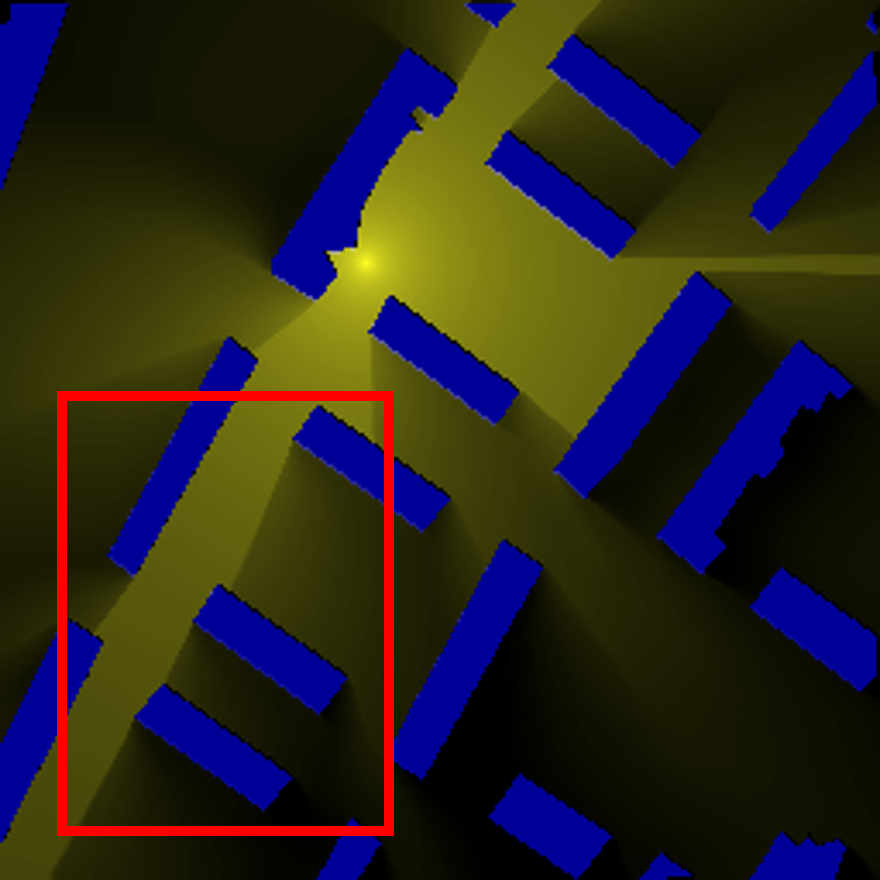} & 
        \includegraphics[width=0.35\columnwidth]{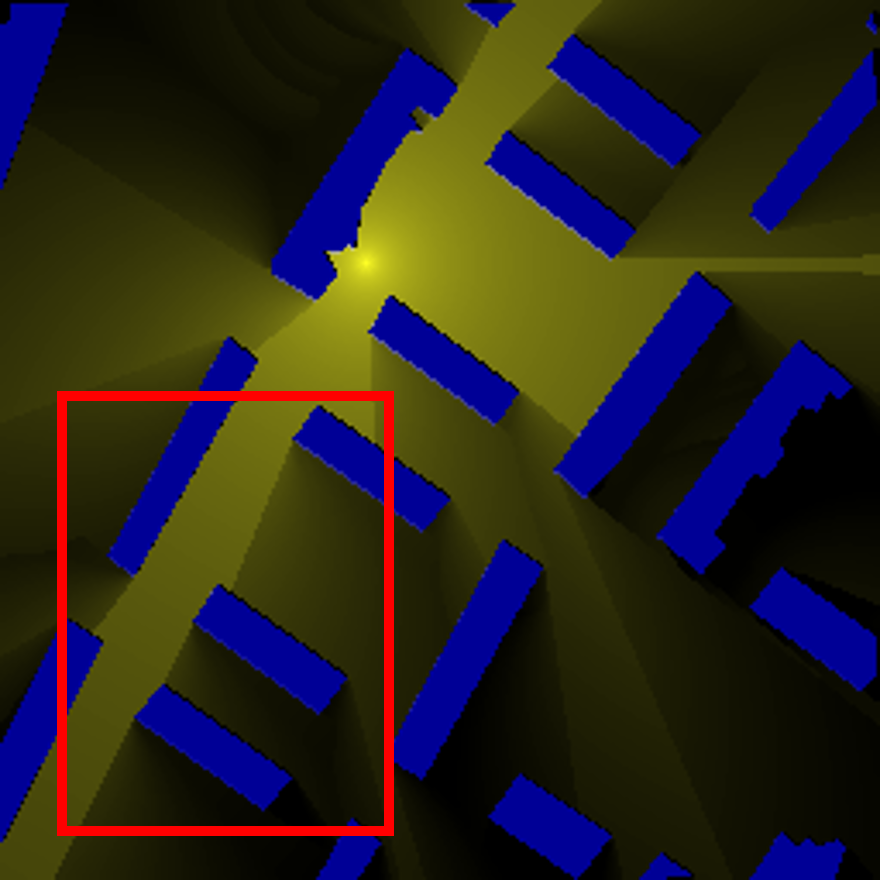} \\
        \small (c) RadioMamba (Full Block) & \small (d) Ground Truth
    \end{tabular}
    \caption{Visual comparison of MambaConvBlock components.}
    \label{fig:ablation_visual}
    \vspace{-12pt}
\end{figure}

The qualitative results, shown in Fig. \ref{fig:ablation_visual}, are illustrative. The Conv-Only model, shown in Fig. \ref{fig:ablation_visual}(a), shows poor performance in predicting pathloss in areas far from the transmitter, producing overly smoothed and blurry constructions. This highlights the inherent limitation of the local receptive field of convolutions for modeling global wave propagation. Conversely, the Mamba-Only model, shown in Fig. \ref{fig:ablation_visual}(b), captures the global structure but lacks fine-grained detail, resulting in blurry shadow edges and an inability to render sharp diffraction patterns. Our full RadioMamba block, shown in Fig. \ref{fig:ablation_visual}(c), combines the strengths of both, producing a map that is both globally coherent and locally precise, closely matching the ground truth shown in Fig. \ref{fig:ablation_visual}(d). This visually confirms that both branches are essential for achieving state-of-the-art results.

\subsubsection{Justification of Depthwise Separable Convolutions}
Our local feature branch utilizes depthwise separable convolutions to maximize efficiency. To quantify this design choice, we trained an alternative model using standard 2D convolutions in the MambaConvBlock. Table \ref{tab:ablation_conv} compares this variant against our proposed model.

Replacing depthwise separable convolutions with standard 2D convolutions did not yield a significant improvement in construction accuracy. The results were nearly identical, with the NMSE changing from 0.0050 to 0.0052. However, this substitution substantially increased the model's complexity, raising the parameter count from 8.6M to 24.71M. This finding confirms that depthwise separable convolutions offer a more effective balance, achieving comparable performance with a considerably smaller model size.

\begin{table}[ht]
    \centering
    \captionsetup{font=small}
    \caption{Comparison of Standard vs. Depthwise Separable Convolutions.}
    \label{tab:ablation_conv}
    \resizebox{\columnwidth}{!}{%
        \renewcommand{\arraystretch}{1.4}
        \begin{tabular}{lcccccc} 
            \toprule
            \textbf{Conv Type} & \textbf{NMSE} $\downarrow$ & \textbf{SSIM} $\uparrow$ & \textbf{PSNR} $\uparrow$ & \textbf{Params(M)} $\downarrow$ & \textbf{Time(s)} $\downarrow$ & \textbf{Memory(MB)} $\downarrow$ \\ 
            \midrule
            Standard Conv & 0.0052 & 0.9657 & 34.10 & 24.71 & 0.0318 & 902 \\ 
            Depthwise (Ours) & \ourmodelval{0.0050} & \ourmodelval{0.9673} & \ourmodelval{34.32} & \ourmodelval{8.60} & \ourmodelval{0.0280} & \ourmodelval{808} \\ 
            \bottomrule
        \end{tabular}%
    }
    \vspace{-6pt}
\end{table}

\subsubsection{Impact of Loss Function Components}
\begin{figure}[ht]
    \centering
    \captionsetup{font=small}
    \begin{tabular}{ccc}
        \includegraphics[width=0.27\columnwidth]{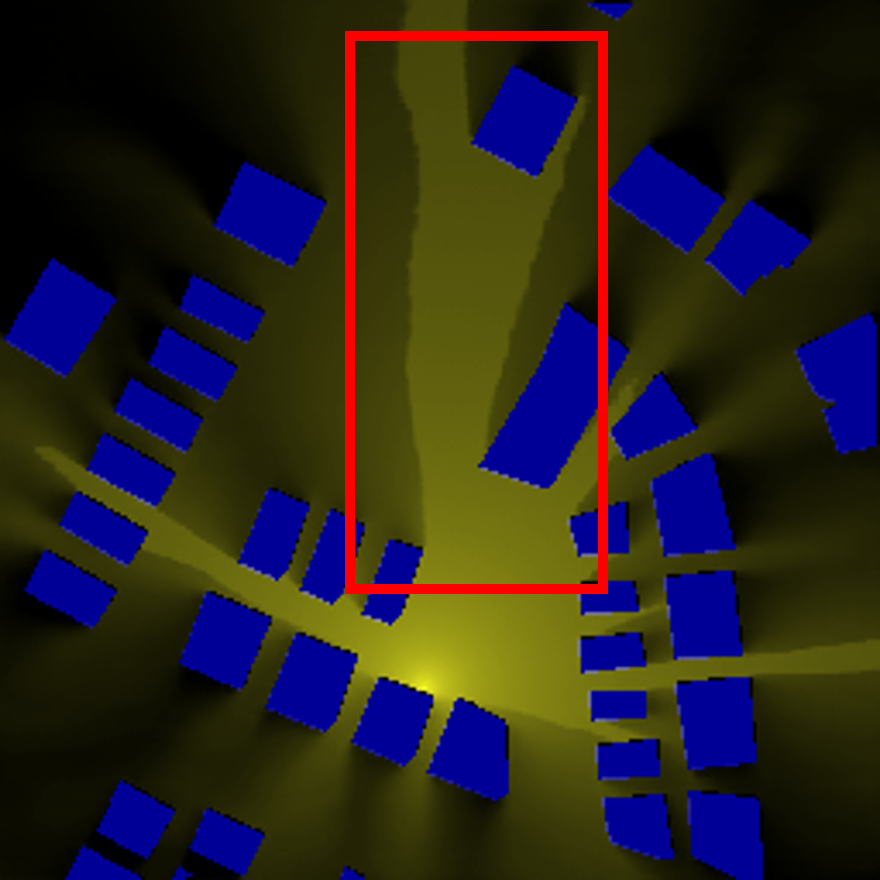} & 
        \includegraphics[width=0.27\columnwidth]{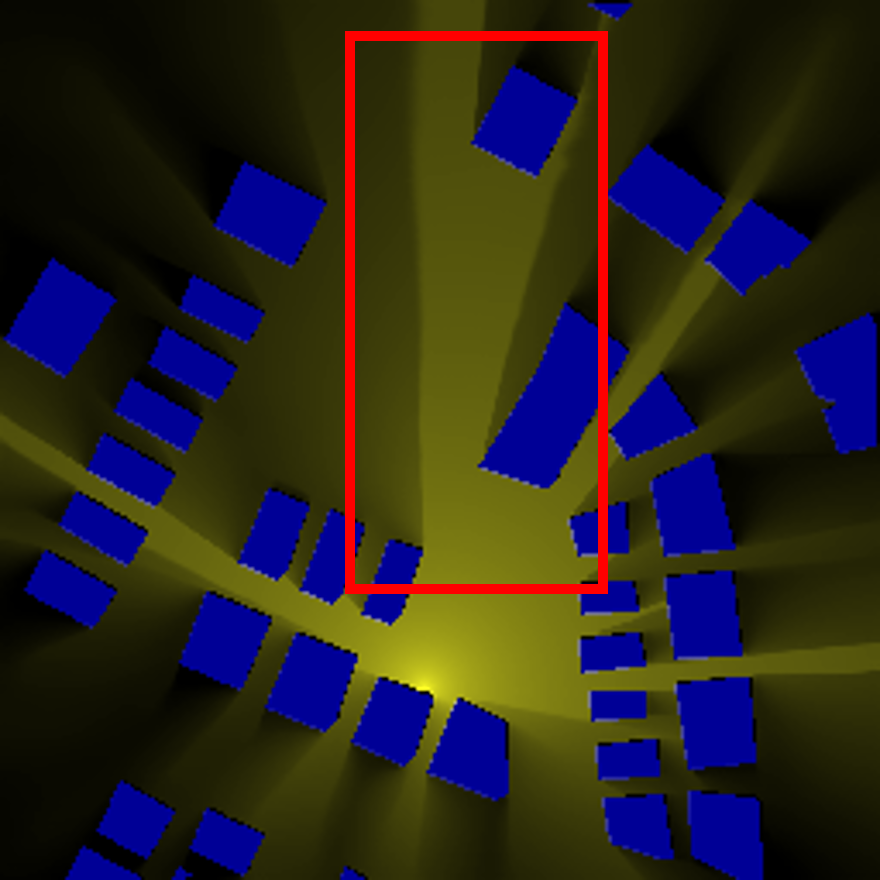} &
        \includegraphics[width=0.27\columnwidth]{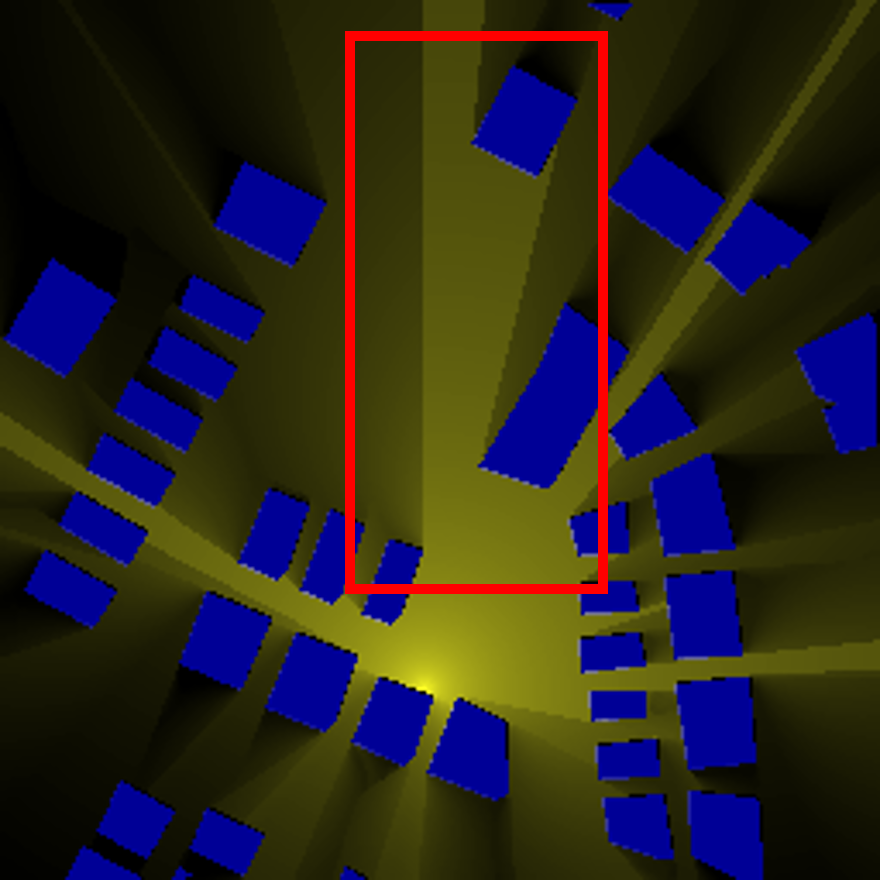} \\
        \small (a) $\mathcal{L}_{L1} + \mathcal{L}_{MSE}$ & \small (b) Full Loss (Ours) & \small (c) Ground Truth
    \end{tabular}
    \caption{Visual impact of the loss function.}
    \label{fig:ablation_loss_visual}
    \vspace{-12pt}
\end{figure}

Finally, we analyze our composite loss function. We trained a model using only the pixel-wise L1 and MSE losses (with weights set to 0.5 each), removing the SSIM and Gradient loss components. Table \ref{tab:ablation_loss} and Fig. \ref{fig:ablation_loss_visual} compare this against our full loss configuration.

The results demonstrate the clear advantage of the composite loss function. The model trained with only L1+MSE loss performs worse across all evaluation metrics. Its NMSE and RMSE are higher, indicating a clear degradation in pixel-level accuracy. Furthermore, the lower SSIM and PSNR scores confirm a tangible loss in perceptual quality and structural fidelity. The visual evidence in Fig. \ref{fig:ablation_loss_visual} corroborates these findings; the map generated with only L1+MSE loss, shown in Fig. \ref{fig:ablation_loss_visual}(a), is structurally inaccurate, with blurry and indistinct shadow edges. In contrast, our full loss function, shown in Fig. \ref{fig:ablation_loss_visual}(b), which includes terms for structural similarity and gradients, produces a map with sharp, physically realistic boundaries that align closely with the ground truth in Fig. \ref{fig:ablation_loss_visual}(c). This confirms that the SSIM and Gradient loss components are not merely for aesthetic improvement; they are essential for guiding the model to learn the correct high-frequency details and structural properties of the radio map, leading to a more accurate and physically plausible solution overall. With all four loss terms retained, we found the current weight ratio to be effective after multiple empirical trials. While a systematic hyperparameter search could potentially yield minor further improvements, we believe the current configuration robustly demonstrates the benefits of our composite loss.

\begin{table}[ht]
    \centering
    \captionsetup{font=small}
    \caption{Impact of Loss Function Components on SRM construction.}
    \label{tab:ablation_loss}
    \resizebox{\columnwidth}{!}{%
        \renewcommand{\arraystretch}{1.4}
        \begin{tabular}{lcccc}
            \toprule
            \textbf{Loss Configuration} & \textbf{NMSE} $\downarrow$ & \textbf{RMSE (dB)} $\downarrow$ & \textbf{SSIM} $\uparrow$ & \textbf{PSNR (dB)} $\uparrow$ \\
            \midrule
            $\mathcal{L}_{L1} + \mathcal{L}_{MSE}$ & 0.0075 & 0.0247 & 0.9541 & 32.39 \\
            Full Loss (Ours) & \ourmodelval{0.0050} & \ourmodelval{0.0199} & \ourmodelval{0.9673} & \ourmodelval{34.32} \\
            \bottomrule
        \end{tabular}%
    }
\end{table}

\section{Conclusion and Future Work}
\label{sec:conclusion}
In this paper, we have introduced RadioMamba, a hybrid Mamba-UNet architecture designed to address the accuracy-efficiency trade-off in radio map construction. Our model's core Mamba-Convolutional block synergistically captures both the global long-range dependencies essential for modeling wave physics and the local features that define environmental boundaries. This design leads to a more comprehensive feature representation, which translates to improved prediction accuracy. Experiments demonstrated that RadioMamba achieves higher accuracy than previous methods, including diffusion models, while being faster and more parameter-efficient. These results suggest that an efficient architectural design aligned with the problem's underlying physics is a viable path to achieving high performance. This balance of accuracy and efficiency makes RadioMamba a suitable candidate for real-time 6G applications by providing the responsive, high-fidelity information needed for network digital twin synchronization or autonomous agent navigation. Future work will focus on extending it to 3D scenarios, investigating on-device deployment, and validating it with real-world data.


\bibliographystyle{IEEEtran}
\bibliography{ref} 

\end{document}